\documentclass[fleqn,usenatbib]{mnras}

\usepackage{newtxtext,newtxmath}

\usepackage[T1]{fontenc}
\usepackage{ae,aecompl}

\usepackage{graphicx}	
\usepackage{amsmath}	
\usepackage{amssymb}	
\usepackage{hyphenat}	
\usepackage{threeparttable}	

\include{Figs}

\newcommand{\kms}{\,km\,s$^{-1}$}	

\newcommand{\CIV}{\ion{C}{iv}}		
\newcommand{\CIII}{\ion{C}{iii}]}	
\newcommand{\MgII}{\ion{Mg}{ii}}	
\newcommand{\SiIII}{\ion{Si}{iii}]}	
\newcommand{\AlIII}{\ion{Al}{iii}}	
\newcommand{\FeII}{\ion{Fe}{ii}}	
\newcommand{\FeIII}{\ion{Fe}{iii}}	
\newcommand{\HeII}{\ion{He}{ii}}   
\newcommand{\SiIV}{\ion{Si}{iv}}   
\newcommand{\OIV}{\ion{O}{iv}}   
\newcommand{\OIII}{\ion{O}{iii}]}   
\newcommand{\NV}{\ion{N}{v}}   

\newcommand{\AIonly}{AI(BI$=$0)}
\newcommand{\AIBI}{AI(BI$>$0)}

\newif\ifproofread
\newcommand{\changemarker}[1]{%
\ifproofread
\textcolor{red}{#1}%
\else
#1%
\fi
}

\newif\ifproofreadii
\newcommand{\changemarkerii}[1]{%
\ifproofreadii
\textcolor{red}{#1}%
\else
#1%
\fi
}

\title[BAL and non-BAL quasars]{BAL and non-BAL quasars: continuum, emission and absorption properties establish a common parent sample}

\author[A. L. Rankine et al.]{
Amy L. Rankine,$^{1}$\thanks{E-mail: alrankine@ast.cam.ac.uk}
Paul C. Hewett,$^{1}$
Manda Banerji$^{1,2}$
and Gordon T. Richards$^{3}$
\\
$^{1}$Institute of Astronomy, University of Cambridge, Madingley Road, Cambridge, CB3 0HA, UK\\
$^{2}$Kavli Institute for Cosmology, University of Cambridge, Madingley Road, Cambridge, CB3 0HA, UK\\
$^{3}$Department of Physics, Drexel University, 32 S.\ 32nd Street, Philadelphia, PA 19104, USA
}

\date{Accepted XXX. Received YYY; in original form ZZZ}

\pubyear{2020}

\begin{document}
\label{firstpage}
\pagerange{\pageref{firstpage}--\pageref{lastpage}}
\maketitle

\begin{abstract}
Using a sample of $\simeq$144\,000 quasars from the Sloan Digital Sky Survey data release 14 we investigate the outflow properties, evident both in absorption and emission, of high-ionization Broad Absorption Line (BAL) and non-BAL quasars with redshifts $1.6 \lesssim z \leq 3.5$ and luminosities $45.3 < \log_{10}(L_\text{bol}) < 48.2$ erg\,s$^{-1}$. Key to the investigation is a continuum and emission-line reconstruction scheme, based on mean-field independent component analysis, that allows the kinematic properties of the \CIV$\lambda$1550 emission line to be compared directly for both non-BAL {\it and} BAL quasars. \CIV\ emission blueshift and equivalent-width (EW) measurements are thus available for both populations. Comparisons of the emission-line and BAL-trough properties reveal strong systematic correlations between the emission and absorption properties. The dependence of quantitative outflow indicators on physical properties such as quasar luminosity and luminosity relative to Eddington-luminosity are also shown to be essentially identical for the BAL and non-BAL populations. There is an absence of BALs in quasars with the hardest spectral energy distributions (SEDs), revealed by the presence of strong \HeII$\lambda$1640 emission, large \CIV$\lambda$1550-emission EW and no measurable blueshift. In the remainder of the \CIV\ emission blueshift versus EW space, BAL and non-BAL quasars are present at all locations; for every BAL-quasar it is possible to identify non-BAL quasars with the same emission-line outflow properties and SED-hardness.
The co-location of BAL and non-BAL quasars as a function of emission-line outflow and physical properties is the key result of our investigation, demonstrating that (high-ionization) BALs 
and non-BALs represent different views of the same underlying quasar population.
\end{abstract}




\section{Introduction}
Quasar-driven outflows are now widely invoked in galaxy formation models in order to reproduce the observed properties of massive galaxies. Broad absorption lines (BALs) observed in the ultraviolet (UV) spectra of quasars are long established as providing evidence for the presence of high-velocity outflows among a substantial fraction of the quasar population \citep{Weymann1991ComparisonsObjects, Hewett2003TheQuasars, Allen2011AFraction}. Of the $\simeq$40 per cent of luminous, optically-selected quasars classified as BAL-quasars \citep{Dai20082MASSBALQSOs, Allen2011AFraction} the majority, so called `HiBALs', show absorption of highly-ionised species such as \CIV\,$\lambda$1549,  \ion{Si}{iv}\,$\lambda$1397, \ion{N}{v}\,$\lambda$1240 and \ion{O}{iv}\,$\lambda$1034. LoBAL-quasars also show absorption of low-ionisation species (e.g., \AlIII\,$\lambda$1857 and \MgII\,$\lambda$2800), while FeLoBALs are quasars which additionally display \FeII\ and \FeIII\ absorption over extended wavelength intervals. The broad, blueshifted absorption is believed to result from the presence of outflowing gas along the direct line-of-sight. Evidence for the presence of infalling or rapidly rotating material from broad absorption lines is rare \citep{Hall2013BroadOutflows}.

Following an original observation by \citet{Gaskell1982AMotions}, increasing attention over the last decade has been devoted to quantifying the outflow-signatures evident from the high-ionisation emission lines in the ultraviolet \citep[e.g.,][]{Sulentic2000EigenvectorNuclei, Leighly2004HubbleInterpretation, Richards2011UnificationEmission} -- particularly the blueshifting of the prominent \CIV$\lambda$1549 emission.

Disc winds \citep{Murray1995AccretionNuclei, Elvis2000AQuasars, Proga2000DynamicsNuclei, Proga2003NumericalForces} are generally considered to explain the origin of both the observed BALs and the blueshifted emission. From a thin accretion disc, a wind emerges as a result of radiation pressure from the disc's UV emission \citep[see fig. 1 of][]{Murray1995AccretionNuclei}. Some models invoke shielding of the wind \citep{Murray1995AccretionNuclei, Proga2000DynamicsNuclei} or perhaps disc geometry \citep{Leighly2004HubbleInterpretation, Luo2015X-RAYDISK} to explain the high velocities that the gas can reach without being over-ionised by the central source. 
Radiation line-driving by ultraviolet photons has been considered as an acceleration mechanism of the wind \changemarkerii{\citep[e.g.,][]{Giustini2019AContext}} and signatures of this process, such as line-locking, have been observed in both narrow absorption lines and BALs \citep{Bowler2014Line-drivenQuasars, Mas-Ribas2019TheOutflows}. \changemarkerii{However, acceleration by magnetohydrodynamics has also been considered \citep{Blandford1982Hydromagneticjets} and it is probable that both acceleration mechanisms are responsible for the observed winds.}

In a number of models the properties of outflowing material are orientation dependent, e.g., relative to the black-hole spin axis, and the probability a quasar is observed to possess BAL features depends on viewing angle. There is, however, still no agreement on whether BALs do arise as the result of viewing angle or represent a particular phase in the fuelling-outflow cycle, with the wind potentially clearing the absorbing gas from the galaxy as the quasar transitions to a non-BAL quasar. With multiple observations of the same object now available, it is possible to observe the variability of the BALs on rest-frame timescales of $\leq$5 years with some objects transitioning from BAL quasars to non-BAL quasars \citep{FilizAk2012BroadSample, FilizAk2013BroadSample, Sameer2019X-rayTransformations} \changemarkerii{and vice versa \citep{Rogerson2018EmergenceOutflows}}. Such transformations may arise due to changes in the ionising state of the outflowing gas or result from wind motion transverse to the line-of-sight \citep[e.g.,][]{Gibson2010TheTimescales, Sameer2019X-rayTransformations}.


The focus of this paper are the far more common HiBAL objects amongst the BAL-quasar population and we exclude consideration of the much smaller fractions of LoBAL and FeLoBAL quasars. Careful comparisons of the emission-line properties of HiBAL and non-BAL quasars have been undertaken \citep[e.g.,][]{Weymann1991ComparisonsObjects, Reichard2003ContinuumQuasars, Baskin2013TheParameters, Baskin2015OnQuasars}. Results of such investigations have shown that, in general, the UV and optical SEDs, unaffected by the presence of BAL troughs, of BAL and non-BAL quasar populations are very similar. Relatively weak emission features, however, which are diagnostics of the far-UV ionising spectrum, such as \HeII\,$\lambda$1640, have been found to show systematic differences between the populations \citep{Baskin2013TheParameters}. To date, the strong absorption present in BAL-quasars has precluded a direct comparison of the kinematic properties of high-ionisation ultraviolet emission lines, such as \CIV\,$\lambda$1549, in the BAL- and non-BAL-quasar populations. As a consequence, investigations of the outflow properties of luminous quasars have been confined to evidence from {\it either} absorber- {\it or} emission-properties.

The key aim of this paper is to provide the first direct comparison of the underlying ultraviolet spectral energy distributions (SEDs) of BAL and non-BAL quasars and thus to investigate outflow properties using {\it both} absorber and emission diagnostics. The ultimate goal is to use the relationships between the absorption- and emission-outflow properties to constrain the geometry and physical parameters of a disc wind or other models. In this first paper, however, we concentrate on presenting the observational results for a large sample of BAL and non-BAL quasars. 

The structure of the paper is as follows. In Section~\ref{sec:problem} we review the task of recovering the unabsorbed ultraviolet SEDs of BAL-quasars and the approach taken to perform the investigation here. The selection of the quasar sample is described in Section~\ref{sec:sample} before the specific recipe for the spectral reconstructions is presented in Section~\ref{sec:specfit}. The question of the definition of the sample of quasars that possess `broad-absorption' is the topic of Section~\ref{sec:BALpop}, where both the classic definition, involving the balnicity-index (BI) from \citet{Weymann1991ComparisonsObjects}, and the more extensive absorption-index (AI) from \citet{Hall2002UnusualSurvey}, are employed. The results of the investigation are presented in Section~\ref{sec:res} before a short discussion of the implications in Section~\ref{sec:disc}. An overview of the main conclusions is included in Section~\ref{sec:conclude}.

Vacuum wavelengths are employed throughout the paper and
we adopt a $\Lambda$CDM cosmology with $h_0=0.71$, $\Omega_{\rm M}=0.27$ and $\Omega_{\Lambda}=0.73$ when calculating quantities such as quasar luminosities.

\section{The problem at hand: reconstructing quasar spectra}
\label{sec:problem}
Performing a statistical analysis of quasar UV spectra to produce a small number of `components', linear combinations of which are capable of reproducing individual quasar SEDs, is long established \citep{Francis1992AnSpectra, Yip2004DistributionsSurvey}. Application of the component-reconstruction techniques has proved successful in the context of BAL quasars, where it is necessary to reconstruct the unabsorbed SED at wavelengths affected by the presence of BAL troughs. The SDSS DR12 BAL catalogue \citep{Paris2017TheRelease} is the most extensive sample. The non-negative matrix factorisation (NMF)-based analysis of the SDSS DR6 spectra \citet{Allen2011AFraction} resulted in very effective reconstructions, albeit over a restricted range in wavelength including the \CIV$\lambda$1549 emission line. Implementations involving principal component analysis (PCA) now form key elements of the redshift determination for quasars and the identification of BALs in recent large-scale surveys, particularly the SDSS \citep{Paris2017TheRelease, Abolfathi2018TheExperiment, Paris2018TheRelease}.

From previous work, it is clear that the accuracy of any component-based reconstructions of quasar spectra needs to be high in order to parametrize differences in the emission-line SEDs such as the \HeII$\lambda$1640 properties, particularly for spectra with extensive wavelength ranges affected by BAL troughs. In order to undertake such a quantitative investigation of the underlying UV SEDs of high-ionisation BAL quasars and non-BAL quasars using the very large quasar spectra samples available from the SDSS \citep{Paris2018TheRelease} we have revisited the question of how to obtain high-accuracy reconstructions of individual BAL and non-BAL spectra that possess extensive ranges in SED properties.

The PCA, NMF and mean-field independent component analysis (MFICA) (see below) approaches all involve reconstructions based on linear combinations of derived components. There is, however, a significant variation in the form of quasar UV SEDs due to multiplicative changes as a function of wavelength; the effect of extinction due to dust along the line-of-sight is a familiar example. Such intra-population non-linear variations are not optimally characterized via PCA/NMF/MFICA analyses and we have therefore applied an empirical large-wavelength scale correction to the overall shape of the quasar spectra prior to deriving components and generating reconstructions of individual spectra. Essentially, the individual spectra are given the same overall shape and we term the process `morphing' throughout the paper. The application of the large-scale multiplicative shape-morphing has the advantage of reducing the number of components required to produce a reconstruction of specified accuracy. A full description of the procedure is given in Section~\ref{sec:specfit}.

To generate the components that allow spectrum reconstructions we have used MFICA decompositions \citep{Hojen-Sorenson2002Mean-FieldAnalysis, Opper2005ExpectationInference, Allen2013ClassificationAnalysis}. For a specified number of components, the MFICA works as well or better than any decomposition scheme we have investigated (Allen \& Hewett, in preparation). As a result, relatively few components are required to produce reconstructions, making reconstructions of only partially complete spectra (as required for the BAL quasars) more stable.

The question of reconstruction reliability and stability for spectra with extensive missing regions, such as the underlying emission spectra for BAL quasars with large troughs, is key and discussed in some detail by \citet{Allen2011AFraction}. In our analysis we are able to place priors on the component weights when reconstructing wavelength regions affected by BALs (particularly the \CIV\,$\lambda$1549 emission and wavelength regions to the blue) using the properties of the individual quasar spectra unaffected by BAL troughs. In more detail, our scheme uses priors on the MFICA component-weights based on the properties of the \CIII$\lambda$1908-\SiIII$\lambda$1892-\AlIII$\lambda$1857 complex, which are closely related to the morphology of the emission in the 1200--1600\AA\ range as evident in fig. 11 of \citet{Richards2011UnificationEmission}.

\begin{figure*}
    \centering
    \includegraphics[width=\linewidth]{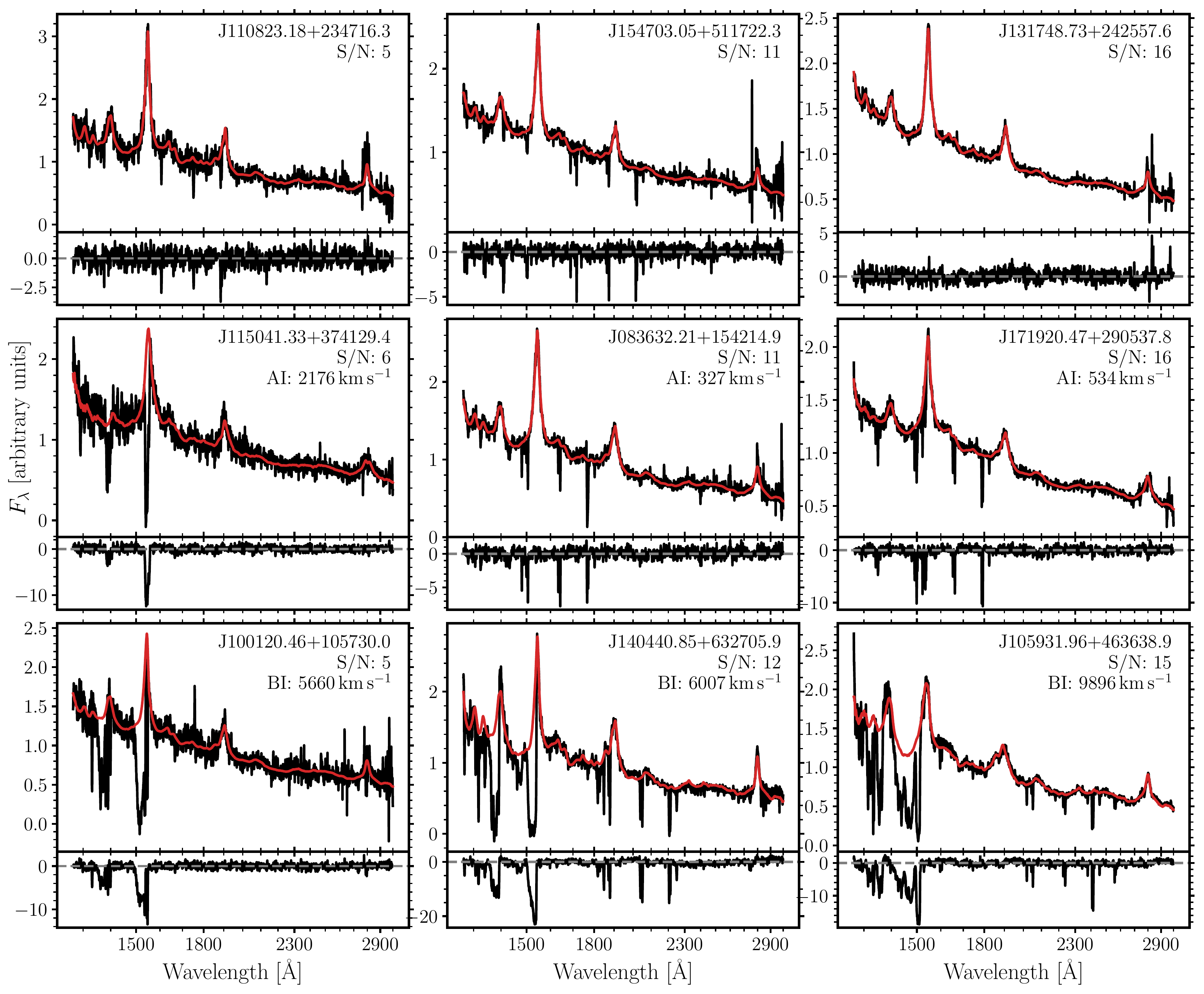}
    \caption{Example spectra (black; smoothed by 5 pixels) and reconstructions (red) showing the quality of the reconstructions across the full range of spectrum S/N. Also plotted, are the residuals (spectrum $-$ reconstruction) divided by the spectrum noise at each pixel. Non-BAL quasars are plotted in the top row and classically-defined BAL quasars are in the bottom row. The middle row contains spectra with absorption too narrow to have positive balnicity index (BI) but with positive absorption index (AI). The spectrum S/N increases from left to right.}
    \label{fig:examp}
\end{figure*}

Using the MFICA-component scheme, we have the ability to reconstruct 95 per cent of (non-) BAL quasar spectra to an accuracy\footnote{\changemarkerii{Defined as the average (spectrum $-$ reconstruction) residuals divided by the spectrum noise at each pixel.}} of 93 (94) per cent over wavelength ranges 1265--3000\,\AA. The procedure is described in detail in Sections~\ref{sec:specfit}-\ref{sec:BALpop} but readers interested primarily in the results may wish to inspect \autoref{fig:examp}, which shows the quality of the reconstructions achieved, and move on to Section~\ref{sec:res}.

\section{Quasar Sample}
\label{sec:sample}

The quasar sample is based on the catalogue compiled from the fourteenth data release of the Sloan Digital Sky Survey \citep[SDSS DR14Q; ][]{Paris2018TheRelease}, which contains $\simeq$526\,000 quasars. The Baryon Oscillation Spectroscopic Survey (BOSS) involved observations of galaxies as well as quasars to allow an investigation of the intervening Lyman-$\alpha$ forest absorption. Unlike the initial SDSS survey, which resulted in the \citet{Schneider2010TheRelease} quasar catalogue drawn from the SDSS DR7, determination of a quantitative selection-function for quasars as a function of SED, redshift and magnitude was not attempted for the BOSS survey. As a consequence, completeness information for quasars \citep{Richards2002SpectroscopicSample} and BAL-quasars \citep{Allen2011AFraction} is not available. The situation is compounded by the fainter magnitude limit to which SDSS DR14 extends and the significant fraction of quasar spectra with very low signal-to-noise ratio (S/N).

Fortunately, for the comparative investigation presented here, it is not necessary to determine the detection probability $P_{qso}$(SED,$m$,$z$) but only to ensure that the probabilities for BAL and non-BAL quasars are similar. The detectability of BAL-troughs decreases significantly as the spectrum continuum S/N decreases \citep{Allen2011AFraction}. Only the most extreme BAL-quasars are, therefore, identifiable in the DR14 quasars that possess low continuum S/N. As the intention is to investigate the relationship between BAL-quasar absorption properties and their underlying emission-line properties, a minimum spectrum S/N-threshold for inclusion of quasars is adopted. 

\begin{figure}
    \centering
    \includegraphics[width=\linewidth]{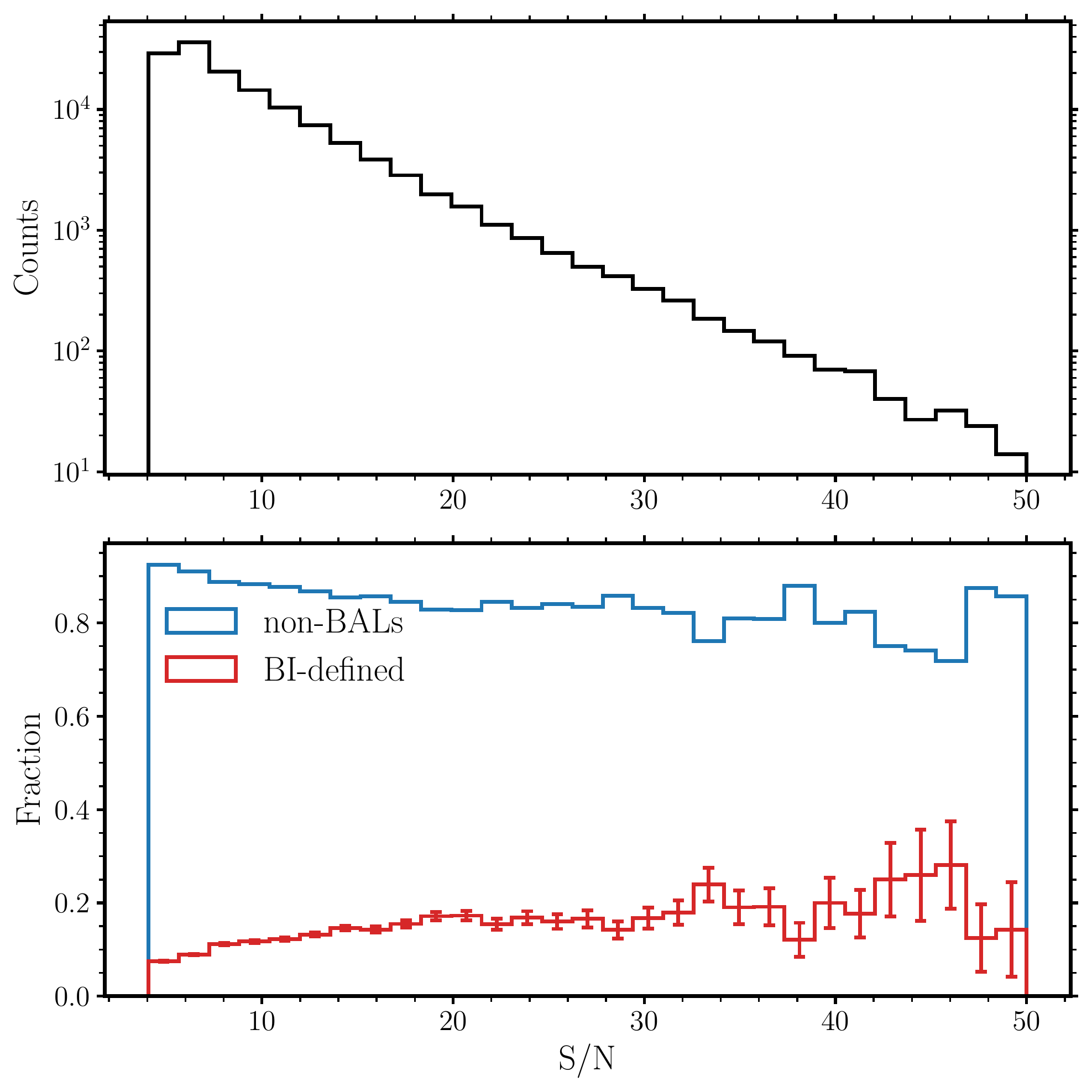}
    \caption{Top: Distribution of average S/N for our quasar sample. Bottom: fraction of BAL and non-BAL quasars as a function of S/N with Poisson errors on the BAL fraction. At each S/N bin, the fraction of BALs and non-BALs adds to unity. Note the drop in BAL quasars at low-S/N is due to the difficulty in detecting troughs in noisy spectra.}
    \label{fig:SN}
\end{figure}

\changemarkerii{The majority of spectra included in the sample possess an average S/N$\ge$5.0 threshold, per SDSS-spectrum pixel, over the rest-frame wavelength interval 1600--2900\,\AA \ (when \MgII $\lambda$2800 is present in the spectrum) or 1600--2000\,\AA \ (at higher redshifts)\footnote{For $\simeq$6\,per cent of the sample the S/N$\ge$5.0 threshold applies to more restricted wavelength intervals that include the low-ionisation \MgII\ and \CIII-emission complex but none of the conclusions of the paper depend on the inclusion/exclusion of the small percentage of such spectra.}. \autoref{fig:SN} plots the fraction of BAL and non-BAL quasars as a function of average S/N. As expected, the fraction of BAL quasars decreases at low-S/N.}

The classification of quasars as HiBALs or non-BALs is based on the \CIV$\lambda\lambda$1548,1550 emission line and the 25\,000\kms \ region blueward of the line (Section~\ref{sec:BALpop}). Minimum redshifts of $z=1.56$ for quasars with SDSS-III spectra (minimum $\lambda_{obs}\simeq 3600$\,\AA) and $z=1.67$ for quasars that possess only spectra from DR7 (minimum $\lambda_{obs}\simeq 3800$\,\AA) are therefore adopted. The number of quasars with redshifts $z>3.5$ whose spectra satisfy the S/N-threshold is small and a maximum redshift of $z=3.5$ is also imposed. The sample of quasars satisfying the redshift-interval constraints and the spectrum S/N-threshold numbers \changemarkerii{$\simeq$144\,000. Our sample contains only the default spectra listed in DR14Q of quasars for which duplicate observations exist.}

The estimation of systemic redshifts for quasars in the SDSS data releases has improved significantly since DR7 \citep{Schneider2010TheRelease} and the reanalysis of \citet{Hewett2010ImprovedSpectra}. Extensive discussion of the effectiveness of the improved schemes is contained in \citet{Paris2017TheRelease, Paris2018TheRelease} and the implications for clustering investigations are reviewed by \citet{Zarrouk2018The2.2}. Notwithstanding the improvements, our own investigations demonstrate that substantial advances in the accuracy of systemic redshift estimates can be made relative to those included in \citet{Paris2018TheRelease}. 

Here, quasar redshifts are calculated using spectrum reconstructions based on a mean-field independent component analysis scheme \citep{Allen2013ClassificationAnalysis} with the redshift as a free parameter. The reconstructions are deliberately confined to the 1600--3000\,\AA \ region, \textit{thereby excluding the C\,{\small IV}-emission line}. The stability of the rest-frame wavelengths of the low-ionisation emission lines \ion{O}{I}$\lambda$1304+\ion{Si}{II}$\lambda$1307, \ion{C}{II}$\lambda$1335, \AlIII$\lambda$1857, \SiIII$\lambda$1892, \CIII$\lambda$1908 and \MgII$\lambda$2800, independent of the large range of ultraviolet SEDs, including high-ionisation emission-line blueshifts, is used to verify the effectiveness of the redshift determinations. In terms of the quasar rest-frame velocity differences, in excess of 35\,000 quasars show a shift of $>$500\kms between the DR14 and our new redshifts, with $\simeq$9800 quasars possessing shifts of $>$1000\kms. While the use of the new redshifts has an impact on the effectiveness of the spectral reconstructions and which objects are classified as BAL-quasars (Section~\ref{sec:BALpop}), none of the statistical results presented in Section~\ref{sec:res} change if the DR14 redshifts are employed. 

\begin{figure}
    \centering
    \includegraphics[width=\linewidth]{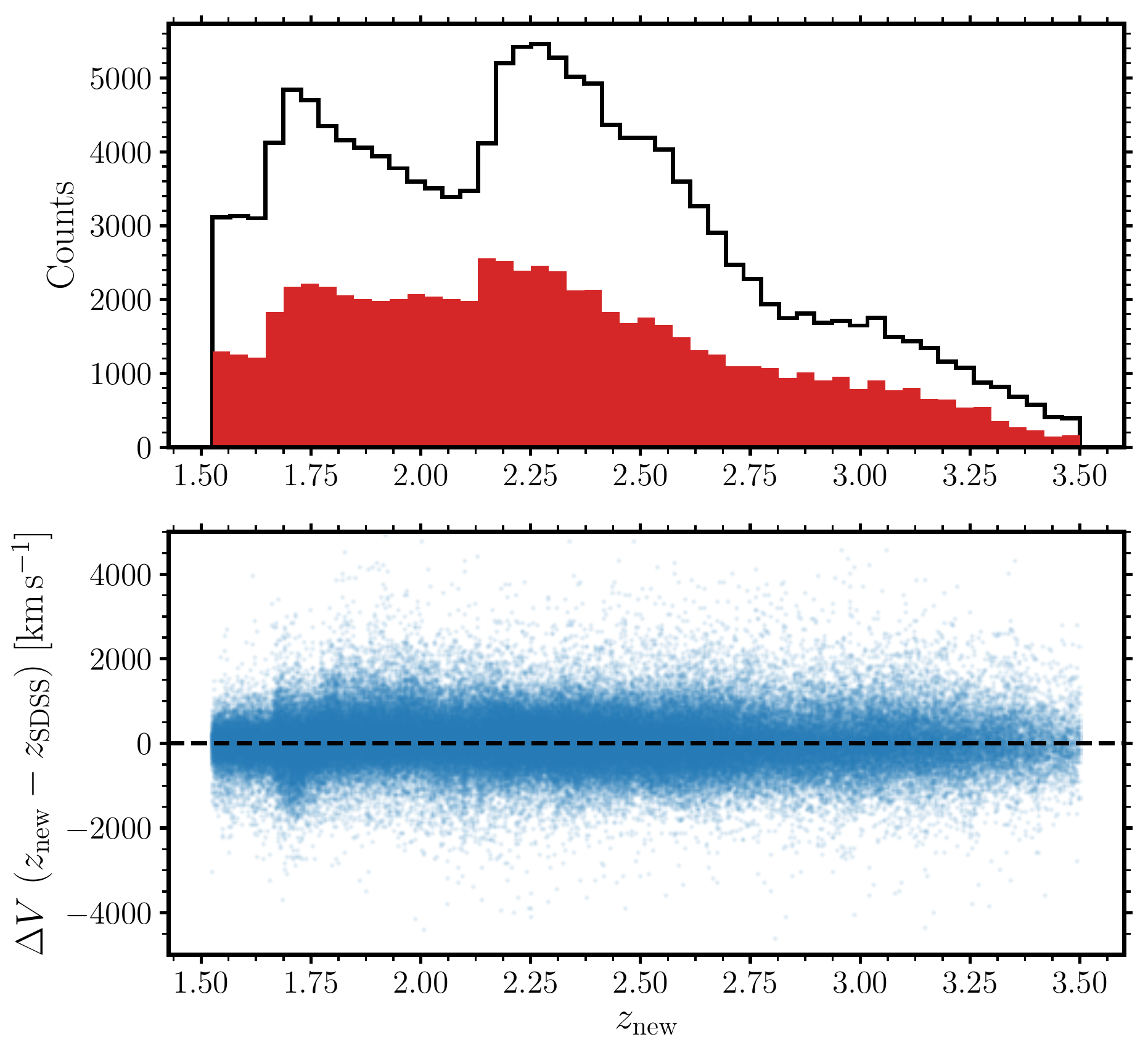}
    \caption{Comparison of DR14Q redshifts, $z_{\text{SDSS}}$, with our new redshifts, $z_{\text{new}}$. Top: distribution of corrected redshifts of all quasars in our base sample (black) and the sub-sample of classically-defined BAL quasars (red; see Section~\ref{sec:BALpop}). The BAL quasar counts have been multiplied by four to better show the $n(z)$-distribution. Bottom: difference between $z_{\text{new}}$ and $z_{\text{SDSS}}$, as a velocity in the rest-frame, as a function of $z_{\text{new}}$.}
    \label{fig:DR14_z}
\end{figure}

The redshift distribution of the base 143\,664-quasar sample to be analysed is presented in \autoref{fig:DR14_z} along with a comparison of the SDSS DR14 redshifts with our new values. 
The first stage in the analysis is to generate reconstructions for each of the quasar spectra as described in the next section.

\section{Quasar Spectra Reconstructions}
\label{sec:specfit}

The rest-frame wavelength range selected for the investigation is 1260--3000\,\AA. The long-wavelength limit ensures the inclusion of the \MgII$\lambda$2800 emission line for quasars with redshifts up to $z\simeq 2.4$ in the DR14 spectra. The short-wavelength limit allows BAL troughs associated with the \OIV+\SiIV \ emission at 1400\,\AA \ to be included as well as emission due to the low-ionisation species \ion{O}{I}$\lambda$1304+\ion{Si}{II}$\lambda$1307 and \ion{C}{II}$\lambda$1335. The strong BAL absorption associated with the \NV$\lambda$1242 emission and Lyman-$\alpha$ emission are in principle also of interest. The presence of absorption from the Lyman-$\alpha$ forest, however, complicates the problem of reconstructing the intrinsic quasar SEDs, particularly where the aim is to utilise only a small number of components to achieve the reconstructions. 

\subsection{Quasar spectrum shape standardisation: `morphing'}
\label{sec:morph}

Unlike the original SDSS DR7 quasar spectra the observational constraints relating to the placement of the fibres for the BOSS DR14 observations mean that differential atmospheric refraction has a significant effect on the spectrophotometry, particularly at blue wavelengths. An extensive discussion of the effect and a scheme for performing post-reduction corrections to the spectrophotometry is presented in \citet{Margala2016ImprovedSample}. 

The differential atmospheric refraction effects give rise to $\pm$10 per cent multiplicative variations in the spectrophotometry. Variations in the amount of dust extinction affecting the ultraviolet continuum and emission-line fluxes from quasar to quasar also produce `intrinsic' multiplicative differences in the large-scale shape of the observed spectra. As heralded in Section~\ref{sec:problem} the presence of such wavelength-dependent multiplicative changes complicates any scheme whereby spectra are reconstructed using linear-combinations of fixed `components'. The first stage in the spectrum reconstruction procedure is therefore to ensure that all the individual quasar spectra possess the same overall shape. The reference SED used in the spectrum-shape morphing is a model quasar spectrum as described in \citet{Maddox2012TheCatalogue}. By construction the model reproduces the ultraviolet SED of luminous, unreddeded quasars and a mean spectrum from the SDSS DR7 quasar catalogue would work equally well. 

\begin{figure*}
\includegraphics[width=\linewidth]{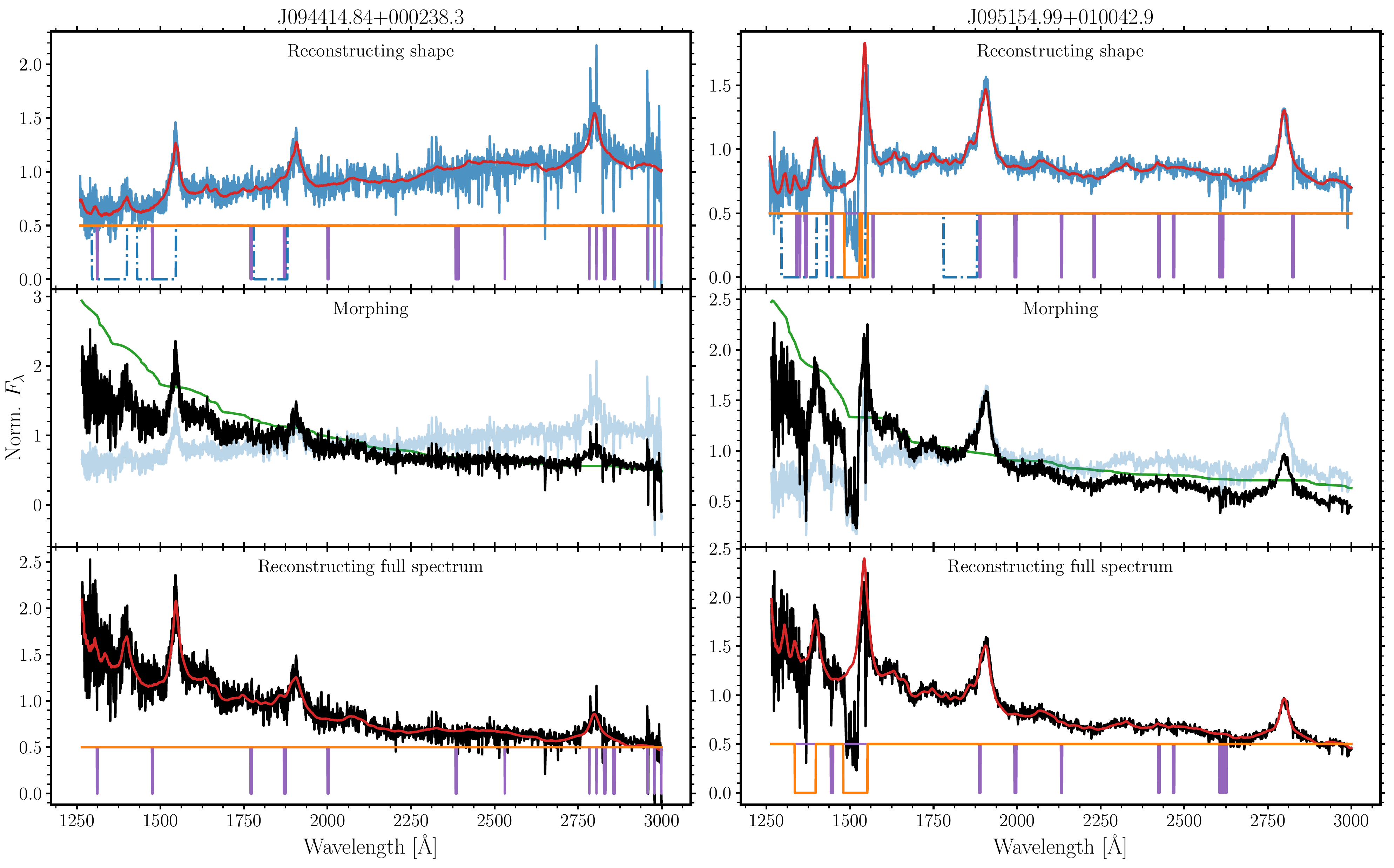}
\caption{A non-BAL quasar spectrum (left) and BAL quasar spectrum (right) illustrating, from top to bottom, the main stages in the MFICA-reconstruction procedure. The two quasar spectra both show significant differences in shape compared to the majority of the quasar population. The original spectra in the top panels (blue) are reconstructed using the BAL-masking routine and the MFICA-components that account for large-scale spectrum shape information. The initial BAL mask (dotted blue) and the iteratively-produced final BAL mask (solid orange) are also shown, along with the narrow absorption-line mask in purple. From the reconstructions in the top panels (red), the morph arrays are produced (green, middle panel; see text for details). The original spectra are multiplied by the morph arrays to produce the black spectra in the middle panels which now have the same overall spectral shape as the master quasar-template spectrum. The morphed-spectra are then fitted with MFICA-derived components to reconstruct the 1260--3000\,\AA\ wavelength interval. The morphed spectra (black), component reconstructions (red) and final BAL- and narrow-absorption-mask (orange and purple respectively) are shown in the bottom panels. It is from such reconstructions that all emission-line and continuum-derived parameters are calculated.}
\label{fig:method}
\end{figure*}

\changemarkerii{The shape-morphing proceeds as follows. Eight MFICA components (Section~\ref{sec:mfica}), to allow spectrum reconstructions, are generated using $\simeq$4000 spectra\footnote{\changemarkerii{MFICA components generated using different samples of $\lesssim$1000 spectra, selected to span the dynamic range in properties of interest (e.g. emission-line equivalent width), are extremely similar. The small-scale, pixel-to-pixel, `noise' in the components does, however, reduce as the number of spectra employed is increased. The DR14-quasar sample is large and all components generated for the investigation are calculated using $\simeq$4000 spectra, with S/N$\geq$8 per pixel and very few masked/bad pixels. The components, and therefore the conclusions, would not change significantly if different, appropriately chosen, samples of 4000 spectra were used.}} of non-BAL quasars with a range of spectrum shapes which have {\it not} been shape-morphed. Individual spectrum reconstructions are then calculated using the masking procedure (to exclude wavelengths affected by both broad and narrow absorption features) described in Section~\ref{sec:fitting} for {\it all} spectra. The scheme thus allows spectra with high-equivalent width BAL troughs to be shape-morphed while ensuring that all spectra are processed in the same way. The resulting reconstructions capture the overall large-scale shape of the spectra.
The continuum in the reference model and each individual reconstruction is then interpolated across the locations of the strongest emission lines. The ratio of the reconstruction continuum and reference model is calculated and median filtered using a 601-pixel ($\simeq$250\,\AA) window. The original spectrum is then multiplied by the resulting smooth function (see top two rows of \autoref{fig:method} for examples of the morphing procedure).
$\sim$76 per cent of spectra are modified by factors of less than $\pm$10 per cent over the full wavelength range.}

\subsection{MFICA-component generation}
\label{sec:mfica}
Samples of morphed spectra are used to calculate the components to be employed in the definitive spectrum reconstructions. First, bad pixels identified in the SDSS reduction pipeline are masked along with strong sky lines (e.g., the 5578\,\AA\ [\ion{O}{i}] emission) in the blue. Narrow absorption features are also masked by applying a 61-pixel median filter to define a pseudo-continuum and masking the pixels where the spectrum lies below this pseudo-continuum by at least 3$\sigma$, where $\sigma$ is the spectrum noise. Three pixels blueward and redward of each such pixel are also masked.

The MFICA-components are calculated \citep[see][]{Allen2013ClassificationAnalysis} using a sample of $\simeq$4000 quasar spectra with complete coverage of the 1260--3000\,\AA \ wavelength range\footnote{The quasar spectra have redshifts in the interval $z\simeq2.0-2.3$.}  and without any BAL-troughs. \changemarkerii{Linear combinations of the resulting 10-components reproduce $\simeq$99 per cent of the non-BAL quasars, with average fractional errors, (spectrum-reconstruction)/reconstruction, less than 2 per cent.} Systematic deviations are greatest for quasars with the most extreme \CIV\ emission equivalent widths. For the purpose here, the form of the MFICA-components is not important, the goal is simply to reproduce the quasar SEDs accurately. Two further sets of components were therefore generated using $\simeq$2000 spectra with \CIV(EW)$>$40\,\AA \ (7 components\footnote{The number of components required is fewer than that for the initial set because of the reduced dynamic range in SED-properties.}) and \changemarkerii{$\simeq$2000 more with }\CIV(EW)$<20$\,\AA \ (10 components) specifically to allow more accurate reconstructions of spectra with extreme emission-line properties. \changemarkerii{Each quasar with weak or strong \CIV(EW) -- measured from an initial reconstruction using the standard set of components -- is fitted with the corresponding extreme-\CIV components and from the standard and the extreme reconstructions, the best is chosen based on a $\chi^2$ measurement.}

\subsection{Fitting the whole quasar sample}
\label{sec:fitting}
The routine for fitting the MFICA components utilises the Python package \nohyphens{\textsc{lmfit}} \citep{Newville2014LMFITPython} for non-linear least squares minimisation using the Levenberg-Marquardt algorithm with priors (\nohyphens{\textsc{L-BFGS-B}}). The priors adopted for the component weights are based on the strong correlation between the morphology of the \CIII-complex and that of the \CIV\ line (see Appendix~\ref{app:priors} for details). Reconstruction of the intrinsic \CIV\ emission when much of it has been absorbed or masked is greatly improved by the use of the priors on the component weights (see discussion below). \autoref{fig:method} illustrates the full fitting procedure.

In addition to the shape morphing and masking of the narrow absorption lines (NALs), in order to successfully fit BAL quasars with the MFICA components, it is crucial that the wavelength regions of the spectra that are affected by BAL troughs are masked. For HiBALs, these are regions blueward of 1400\,\AA\ for \ion{Si}{iv}+\ion{O}{iv} and 1550\,\AA\ for \CIV. Troughs differ greatly in velocity width and position (up to thousands of~\kms) across BAL quasar spectra and to avoid masking more or less of each spectrum than necessary an optimal mask for each spectrum was defined in a similar manner to \citet{Allen2011AFraction}. The routine also takes into account absorption redward of the BAL regions presented above resulting from errors in redshift. The steps applied to the morphed spectra, with flux $F_\lambda$ and noise $\sigma$, are as follows:
\begin{enumerate}
\item \label{itm:init}The initial mask covers regions of the spectra just blueward of the peak of the emission lines where troughs are likely to appear: 1295--1400\,\AA\ for \ion{Si}{iv}+\ion{O}{iv}, 1430--1546\,\AA\ for \CIV\ and 1780--1880\,\AA\ for \ion{Al}{iii}.
\item \label{itm:comps} The components are then fitted to the unmasked parts of the spectrum. 
\item A new mask is created by considering every pixel in turn. For pixel \textit{cpix}, if the majority of the pixels in [\textit{cpix}$-30$,\textit{cpix}$+30$] have $\left(\textrm{reconstruction} - \textrm{spectrum}\ F_{\lambda}\right)> N\sigma$ then pixel \textit{cpix} is masked. Initially, $N=2$.
\item \label{itm:10pix}Once the pixels have been identified, an extra 10 pixels ($\sim$4\,\AA\ at 1550\,\AA) are masked blueward and redward of each of the masked regions.
\item The BAL troughs associated with the \SiIV+\ion{O}{iv}$\lambda$1400 emission are weaker than those associated with \CIV$\lambda$1550. For much of the quasar redshift range the troughs lie at wavelengths where the spectrum S/N is lower and the masking algorithm is not always fully effective. To ensure absorption associated with \SiIV+\ion{O}{iv} is masked effectively, the masked region from \CIV\ in velocity-space is applied to the \SiIV\ region, in addition to what is masked in the iterative process.
\end{enumerate}
Steps~\ref{itm:comps} through \ref{itm:10pix} are repeated until the mask converges, normally after two or three iterations. When there is no mask convergence after 10 iterations, the process is started from scratch at step~\ref{itm:init} with $N=N+0.25$ up to $N=4$, at which point the spectrum is flagged as not having converged.

In order to quantify our ability to accurately reconstruct the BAL quasar spectra we performed tests using 7000 non-BAL spectra covering the full observed-range in \CIV\ emission properties. In detail, we masked portions of the non-BAL quasar spectra across wavelength ranges where BAL troughs appear in the BAL quasar spectra matched in \CIV\ emission properties and S/N and then reconstruct the spectra. To quantify the differences between the spectra and the reconstructions we measure the cumulative difference in the \CIV\ line between 1400 and 1600\,\AA\ as a fraction of the \CIV\ EW as measured from the spectrum:


\begin{equation}
    f(EW) = \frac{1}{\text{EW}}\sum{\frac{\left(\text{spectrum}-\text{reconstruction}\right)}{\text{continuum}}\Delta\lambda}.
\end{equation}

When fitted without masking, 97 per cent of the 7000 non-BAL quasar spectra had $f(EW)<10$ per cent and a median $f(EW)$ of only 2 per cent. The investigation confirmed the importance of employing priors on the component weights (see Appendix~\ref{app:priors}). Unsurprisingly, when the reconstructions are relatively poorly constrained, due to extended masked regions or low spectrum S/N, the use of priors for the component weights is essential. \changemarkerii{$f(EW)$ is anti-correlated with S/N but only rises to a median of 6 per cent for S/N\,$\simeq5$.}

When the spectra were masked to simulate BAL quasars but the priors were not implemented, only 78 per cent of the sample had $f(EW)<10$ per cent even though the median $f(EW)$ increased to only 4 per cent. Upon employing the priors to fit the masked non-BAL quasar spectra, the percentage of spectra with $f(EW)<10$ per cent increased to 85 per cent and maintained a low median $f(EW)$ of 4 per cent. Statistically, the average gain of using the priors appears small for the sample as a whole. However, for the relatively small fraction of spectra with extended masks (corresponding to quasars with large BI values) the reconstructions improve significantly. For the actual BAL-spectra, in the iterative process of fitting the components and defining the mask, 
the component weight priors prevents a bias towards narrower BAL masks and the presence of `dips' in the reconstructions where the troughs are present. \changemarker{The use of priors derived from the properties of the \CIII-complex
emission is predicated on the assumption that the relationship between
the \CIV\ emission properties and the \CIII-complex emission is the
same for both BAL and non-BAL quasar populations. While, by
definition, it is not possible to test the assumption directly for
individual BAL-quasars it is possible to perform a powerful test for
the BAL-population as a whole. The results, presented in
Appendix~\ref{app:similar}, demonstrate that the assumption regarding
the similarities of emission-line properties for the BAL and non-BAL
quasar populations is true to very high accuracy.}
Visual inspection of the sparsely occupied regions at the extremes of the \CIV\ emission EW vs blueshift space \changemarkerii{(Section~\ref{sec:res})}, specifically at i) large negative \CIV\ emission blueshift ($<$-1200\kms), ii) small \CIV(EW) (EW$<$10\,\AA) and iii) the region at the bottom-left of the \CIV\ emission space, with relatively high negative \CIV\ emission blueshift and low \CIV(EW) ($\log_{10}\text{\CIV(EW)} < 
-2.3077\times10^{-4}\times\text{\CIV(blueshift)}+1.3231$) reveal a number of intrinsically pathological quasar spectra and some reconstructions that are clearly sub-optimal. \changemarkerii{The number of spectra in all three regions is small: 3027 spectra or 2 per cent of the quasar sample. 2283 of the spectra removed in this way are classed as non-BAL quasars by the trough parametrization code (see Section~\ref{sec:BALpop}) but upon visual inspection include a significant fraction of LoBAL and FeLoBAL objects and their classification (and large, negative blueshift) is often due to insufficient masking. Spectra in these regions are, therefore, excluded from subsequent analysis. A further 189 spectra were excluded because the reconstruction scheme failed to converge due to poor morphing or the reconstructions had reduced-$\chi^2>2$. The excluded spectra were distributed evenly across \CIV\ emission space.}

For the remaining \changemarkerii{140\,448} quasars, each spectrum has the same overall spectral shape and a continuum plus emission line reconstruction is available. An associated mask array contains the wavelengths of pixels where absorption has been identified. Quantification of the absorber properties and the definition of sub-samples of BAL-quasar spectra is now possible.

\section{Defining the BAL quasar population}
\label{sec:BALpop}

The classical definition of a BAL quasar is a quasar with non-zero balnicity index (BI) as defined by \citet{Weymann1991ComparisonsObjects}:
\begin{equation}
    \text{BI} = \int^{25000}_{3000}\left(1-\frac{f(V)}{0.9}\right)C\ \text{d}V
    \label{eqn:BI}
\end{equation}
with $C=1$ when $f(V)<0.9$ continuously for at least 2000\kms, otherwise $C=0$.

\citet{Hall2002UnusualSurvey}, seeking a measure to incorporate a more liberal definition of `absorption', introduced the absorption index (AI) to include absorption closer to the systemic redshift and as narrow as 450\kms. Here, $C=1$ when $f(V)<0.9$ continuously for at least 450\kms.

\begin{equation}
\text{AI} = \int^{25000}_{0}\left(1-\frac{f(V)}{0.9}\right)C\ \text{d}V.
    \label{eqn:AI}
\end{equation}

The DR14Q catalogue contains measurements of BI but the updated redshifts used in this investigation as well as our own reconstructions mean some spectra change classification based on the BI and AI values. We remeasured the BI and AI using Python code written by \citet{Rogerson2019BALparams} with minor modifications, allowing us to use our reconstructions to produce the normalised spectra, $f(V)$, in Equations~\ref{eqn:BI} and \ref{eqn:AI}. Each pixel in the normalised spectra has a value calculated using the inverse-variance weighted flux within a five-pixel window, thereby reducing the effect of single-pixel noise spikes. The code also calculates a variety of individual absorption trough measurements including minimum and maximum ejection velocities and trough depth. Discussion of the additional trough measurements is included in Section~\ref{sec:BALt}.

\begin{figure}
    \centering
    \includegraphics[width=\linewidth]{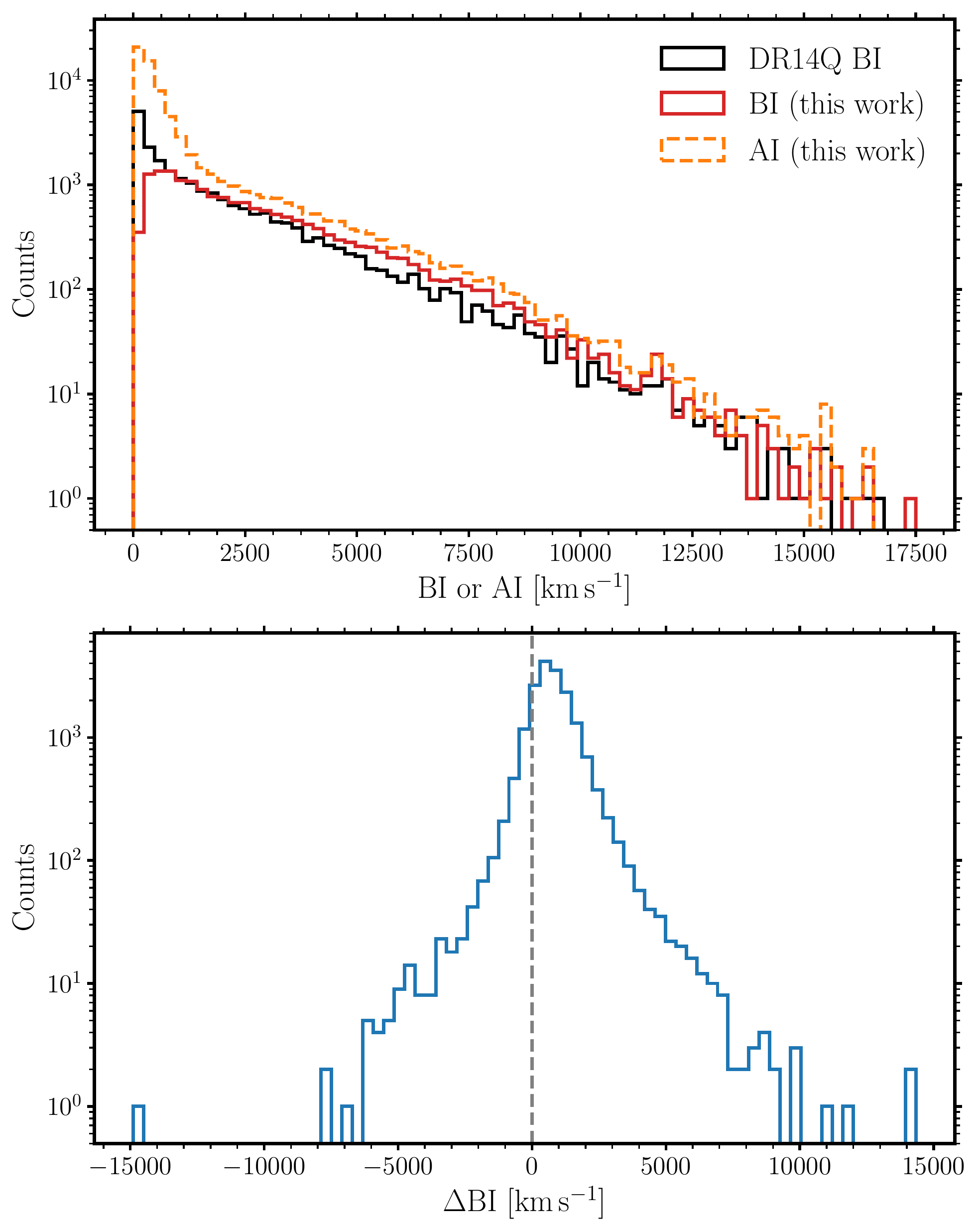}
    \caption{Comparison of the BI$_\text{DR14Q}$ values with our BI values. Top: counts of BI$_\text{DR14Q}$ and BI and AI calculated for this paper. Bottom: $\Delta\text{BI}=\text{BI}-\text{BI}_\text{DR14Q}$. For visualisation purposes, we have removed BI$=$0 or AI$=$0 quasars from each of the histograms. In this work, the calculated BI values are, on average, 700\kms\ larger than those in DR14Q. Below 1000\kms, there is a decrease in the number of BI-defined BAL quasars according to our measurements and compared to the DR14Q values, which we believe is a result of changes in redshift. Above 1000\kms\ there are more BI-defined quasars due to higher continua in our spectrum reconstructions.
    }
    \label{fig:DR14_BI}
\end{figure}

We compare our BI values with those in the DR14Q catalogue in \autoref{fig:DR14_BI}. The redshift changes have primarily affected narrow troughs near to the minimum (3000\kms) and
maximum (25\,000\kms) velocity boundaries used for the BI measurements; more or less of the absorption is shifted into the BI velocity-window. In particular, if the \CIV\ emission is considered when calculating the redshifts of BAL quasars with \changemarkerii{abrupt onset of absorption in the blue half of the line,} 
these quasars can have systematically higher redshifts in SDSS. Our statistically lower redshifts lead to less of the absorption appearing at velocities $>3000$\kms\ and we can calculate lower or even zero BI.

\begin{figure}
    \centering
    \includegraphics[width=\linewidth]{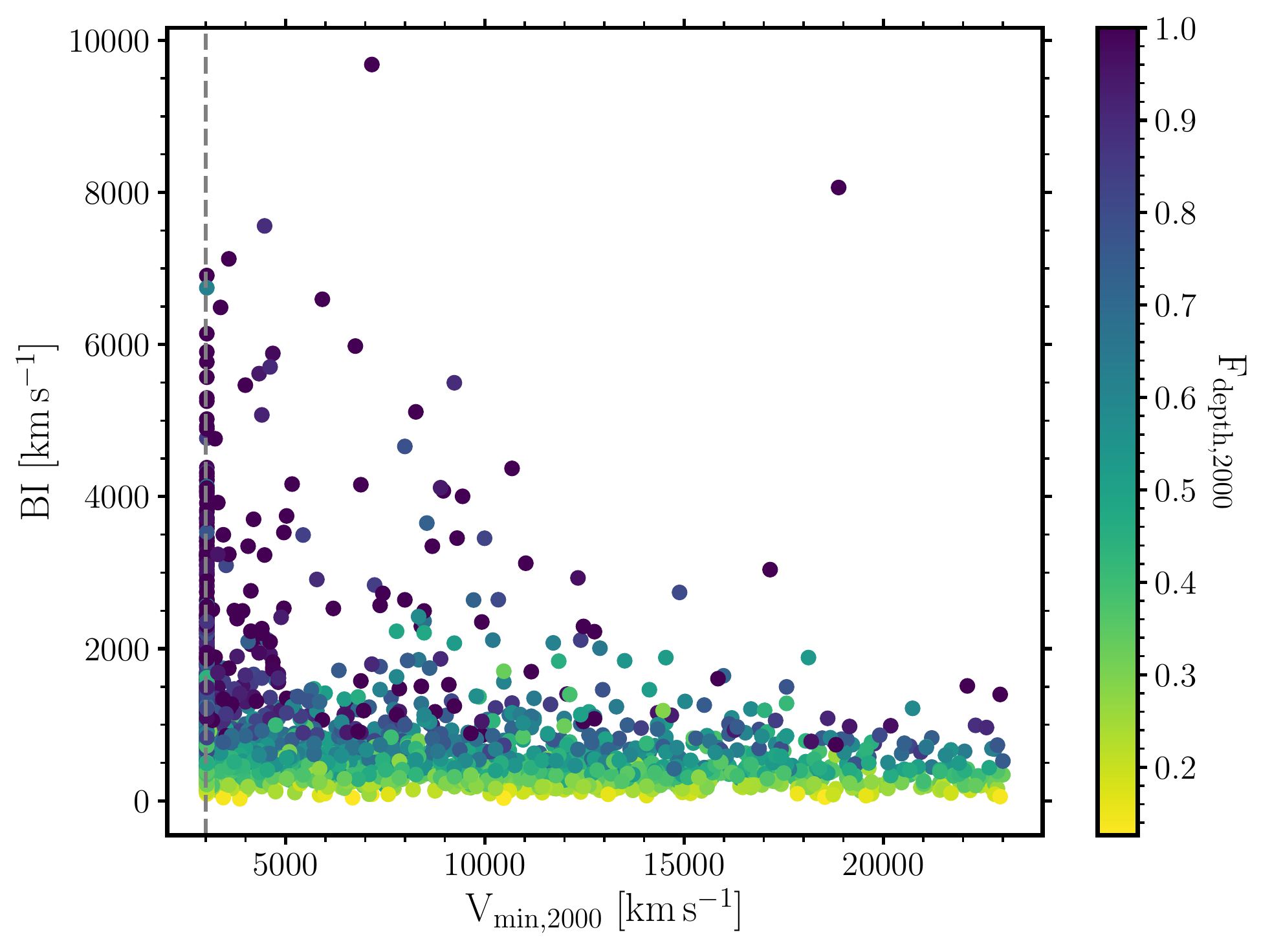}
    \caption{Our BI values against our calculations of the minimum ejection velocity with points coloured by trough depth for the 2359 spectra with BI$>$0 but zero BI$_\text{DR14Q}$. Most of these spectra have minimum velocities close to the 3000\kms\ cut-off where redshift differences are responsible. The spectra with large $V_\text{min}$ have shallow troughs
    and changes in reconstruction due to continuum reconstructions dominate.}
    \label{fig:DR14_B0}
\end{figure}

Differences in the continuum level of the component reconstructions can also result in classification switches. Here, changes result as the contribution of absorption to the BI calculation depends sensitively to the absorption-depth threshold, $f(V)<0.9$ continuously for 2000\kms. In \autoref{fig:DR14_B0} we illustrate these effects by considering the 2359 spectra for which DR14Q listed zero BI whereas we measure BI$>$0. As a result, using the BI-definition we gain 1503 BAL quasars while the 2359 objects in \autoref{fig:DR14_B0} become non-BAL quasars\footnote{While we are confident regarding the use of improved redshifts and reconstructions, none of the results presented in the next section depend on the exact definition of the BAL-quasar sample.}.

We are also able to compare the AI measurements to those in DR12Q BAL catalogue \citep{Paris2017TheRelease}. \citet{Trump2006ARelease} define the reduced-$\chi^2$ for each AI trough in order to remove false-positive troughs which are the result of noise:
\begin{equation}
    \chi^2_\text{trough} = \sum \frac{1}{N} \left(\frac{1-f(V)}{\sigma}\right)^2
\end{equation}
where $N$ is the number of pixels in the trough. We adopt the requirement for $\chi^2_\text{trough}\geq10$ as \citet{Trump2006ARelease} suggest and \citet{Paris2017TheRelease} implement. Differences in the sample of DR12Q AI-defined objects and our sample arise from the changes in redshift and continuum level as for the BI-defined samples. In addition, for the spectra included in DR12Q, our AI-defined sample includes a significantly larger number of quasars. We believe the difference is the result of \citet{Paris2017TheRelease} presenting the AI measurements only for spectra flagged as BAL-quasars through visual inspection. The interpretation is supported by \citet{Paris2017TheRelease} noting that they have an additional $\simeq$27\,000 spectra with non-zero AI not reported in the DR12Q BAL catalogue, which itself contains just $\simeq$22\,000 quasars with AI$>$0.

\changemarkerii{We exclude 1668 LoBAL quasars from our sample (in addition to those in outlying regions of \CIV\ emission space),} defined by a positive BI$_0$ of \AlIII\ troughs where BI$_0$ is a modified balnicity index where the integral in Equation~\ref{eqn:BI} starts at 0\kms. \changemarkerii{An additional 18 FeLoBAL quasars are also removed from the sample by means of the BAL mask (Section~\ref{sec:fitting}): spectra with more than one third of pixels redward of \CIV\ masked are flagged as potential FeLoBAL quasars. The final sample of quasars thus numbers 138\,762, with 73\,806 of those defined as non-BAL quasars, 14\,887 classically-defined BAL quasars (positive BI) and a further 50\,069 with positive AI but a BI of zero.}

\subsection{Two AI populations}
\begin{figure}
    \centering
    \includegraphics[width=\linewidth]{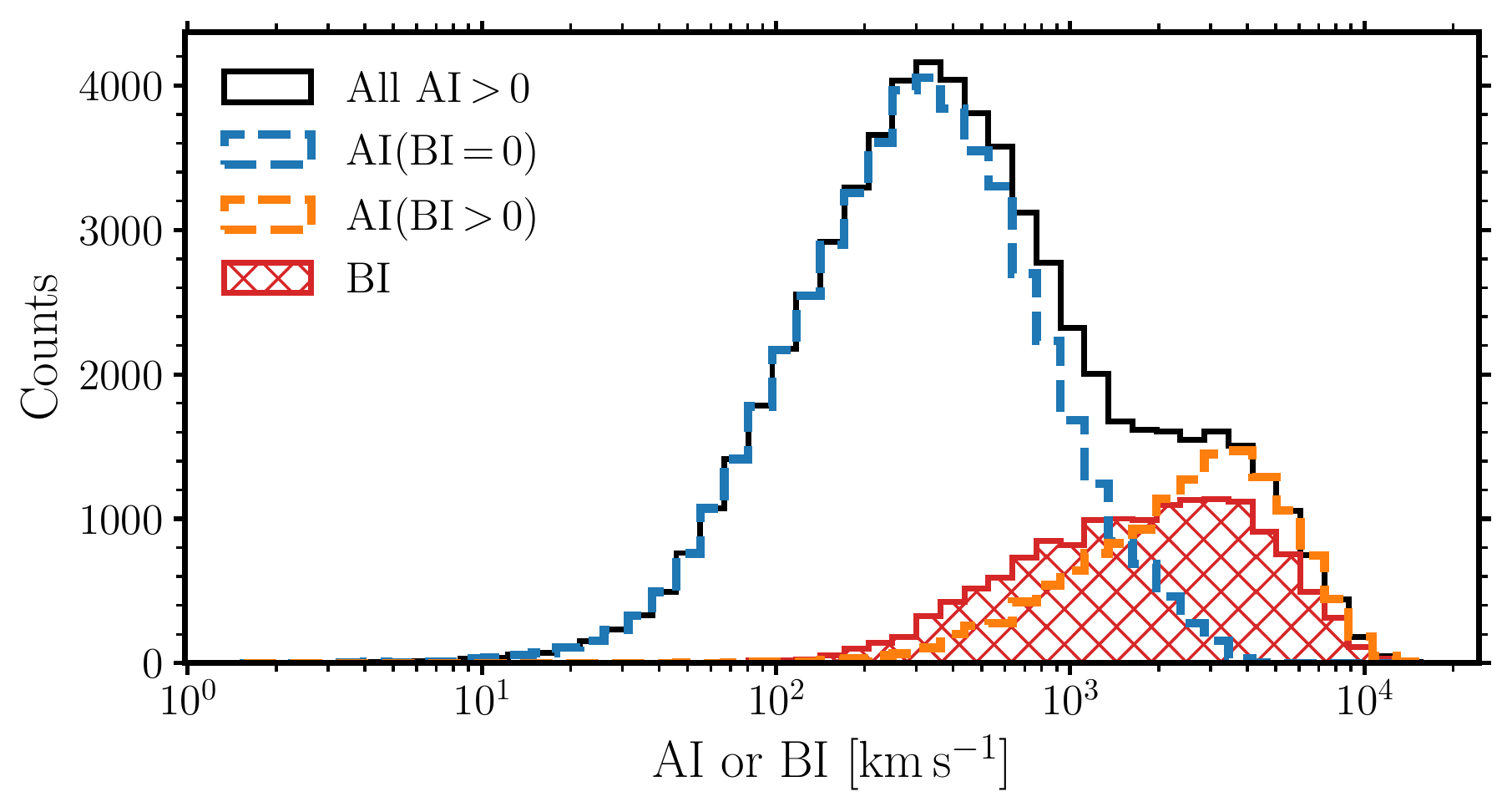}
    \caption{AI distribution for quasars with AI$>$0 (black). The AI population is divided into quasars with zero BI (blue; left column of \autoref{fig:CIVtrough1} where we discuss the trough parameters) and non-zero BI (orange; middle column of \autoref{fig:CIVtrough1}). The form of the AI-distribution suggests strongly that two populations of absorbers contribute. The BI distribution for quasars with BI$>$0 is also plotted (red, hatched; right column of \autoref{fig:CIVtrough1}) which overlaps with the \AIBI\ population since many of the same spectra appear in both populations.}
    \label{fig:AIpops}
\end{figure}

Upon plotting the distribution of AI in log space (\autoref{fig:AIpops}), we observe two populations not unlike \citet[][see their fig. 1]{Knigge2008TheQuasars}, although note that those authors adopt the AI prescription of \citet{Trump2006ARelease} which includes only absorption wider than 1000\kms\ and extends the 25\,000\kms\ maximum velocity to 29\,000\kms. The large-AI population of quasars is dominated by spectra with positive BI (\AIBI) whereas the classically-defined non-BAL quasars dominate the low-AI population (\AIonly). The distribution of BI and \AIBI\ overlap since many of the spectra, thus same trough measurements, appear in both samples. \changemarkerii{The shift of the \AIBI\ toward larger values as compared to the BI values} is a result of the minimum velocity cut-off on the BI. The number of quasars in the \AIBI\ population is also lower than the number in the BI population due to the $\chi^2_{\text{trough}}>10$ requirement of the AI troughs. The properties of the \changemarkerii{\AIonly\ and \AIBI\ }samples are discussed later in Section~\ref{sec:BALt}.

\section{Results}
\label{sec:res}

\subsection
[C IV emission line profile]
{C\,{\sevensize IV} emission line profile}
\label{sec:CIVem}

From the component reconstructions, we can parametrize the \CIV\ emission of both the BAL and non-BAL quasars via a non-parametric approach similar to that of \citet{Coatman2016CIVEstimates, Coatman2017CorrectingMasses}. A power-law continuum $f\left(\lambda\right)\propto\lambda^{-\alpha}$ is fitted to the reconstruction using the median values of $F_{\lambda}$ in the two wavelength regions 1445--1465\,\AA\ and 1700--1705\,\AA. The model continuum is subtracted from the spectrum in the window 1500--1600\AA\ and the emission-line flux computed over the same wavelength interval\footnote{\changemarkerii{For a small number of quasars (1157) with more than 15 per cent of the line flux blueward of the short wavelength limit }
the window is extended down to 1465\,\AA\ to include the majority of the total \CIV\ emission flux.}. The wavelength which bisects the cumulative total line flux, $\lambda_{\text{half}}$, is determined and converted to a velocity to obtain the blueshift of the emission via 
\begin{equation}
    \text{\CIV\ blueshift} = c\times\left(\lambda_r - \lambda_{\text{half}}\right)\mathbin{/}\lambda_r
\end{equation} 
where $c$ is the velocity of light and $\lambda_r$ is the rest-frame wavelength of the emission line, in this instance 1549.48\,\AA\ for the \CIV\ doublet. Positive values correspond to outflowing material towards the observer. The equivalent width (EW) of the emission as well as the location and value of the peak of the line are also calculated. 

\begin{figure}
\includegraphics[width=\linewidth]{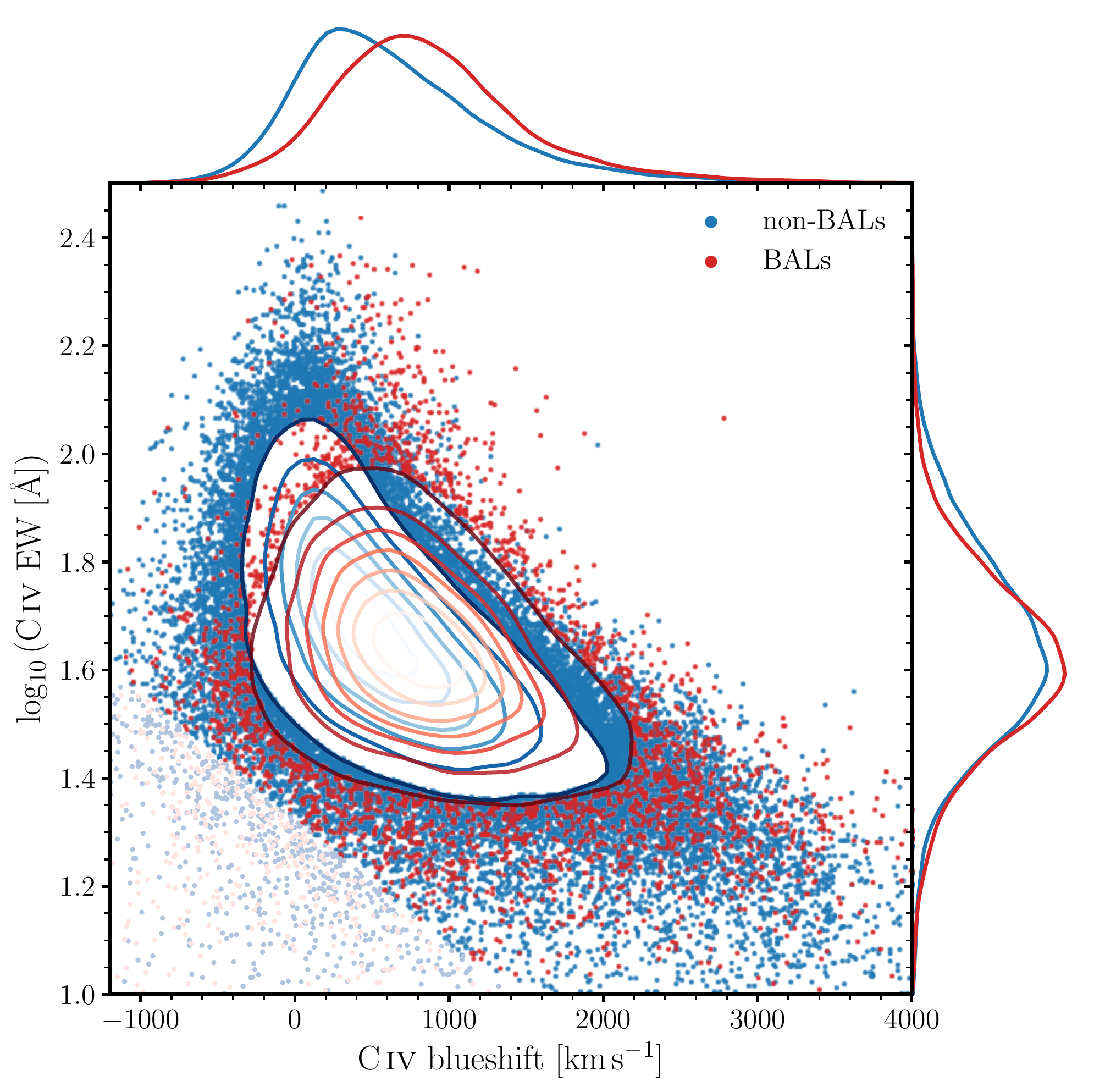}
\caption{\CIV\ emission space for the sample of non-BAL quasars (blue contours/dots) and classically-defined BAL quasars (red contours/dots). Marginalised distributions of blueshift and EW are also shown. The faint points in the sparsely populated lower left of the figure have been excluded from the analysis (Section~\ref{sec:fitting}). The two populations are distributed very similarly within the \CIV\ emission space except at high-EW and low blueshift where there is a notable lack of BALs at the lowest blueshifts.}
\label{fig:CIVnonBALQvBALQ}
\end{figure}

\begin{figure*}
    \centering
    \includegraphics[width=\linewidth]{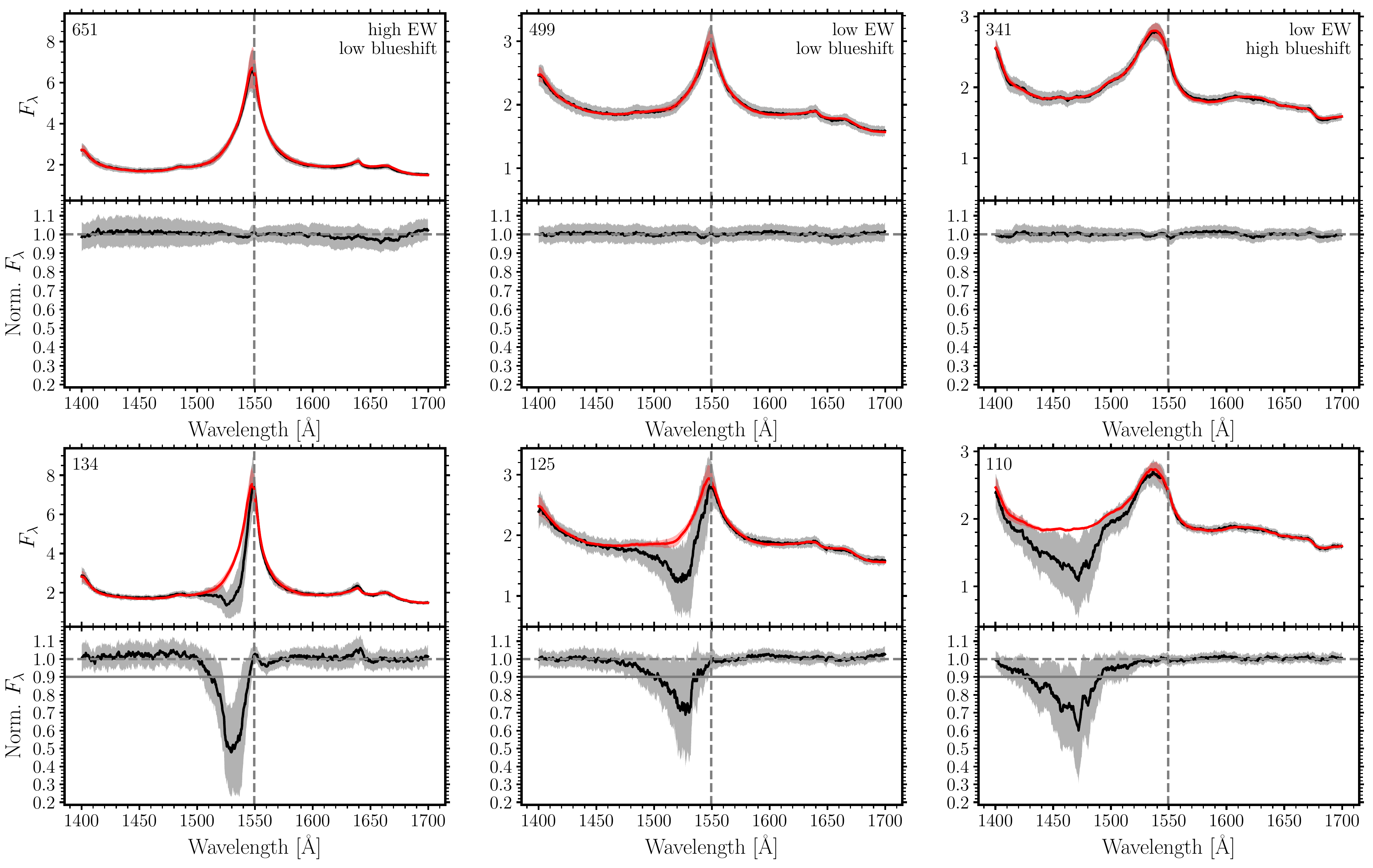}
    \caption{Composite spectra and reconstructions of non-BAL (top) and BAL quasars (bottom) for select regions in \CIV\ emission space. Left: high-EW and low-blueshift \CIV; middle: low-EW and low-blueshift; right: low-EW and high-blueshift \CIV. The number of spectra contributing to each composite is indicated in the top left. In the top panel of each pair, the black line is the median composite SDSS spectrum and the red is the composite reconstruction. The grey and red shaded regions mark the pixel-by-pixel median absolute deviation of the the spectra and reconstructions, respectively. The lower panel of each pair presents the composite spectrum normalised by the composite reconstruction. The shaded region is the median absolute deviation of the normalised spectra. The vertical, dashed, grey line corresponds to the rest wavelength of \CIV\ and the horizontal, solid grey line in the normalised BAL quasar panels marks $f(V)=0.9$, the flux below which BAL troughs contribute to BI and AI calculations. Note that all panels containing the normalised spectra have the same y-scale for comparison. Whilst the \CIV\ emission profiles vary in \CIV\ emission space by design, the BAL troughs also vary systemically. Composite non-BAL and BAL spectra are available to view in an interactive plot of the \CIV\ emission space$^{\ref{foot:url}}$.}
    \label{fig:bal_nbal_compos}
\end{figure*}

\autoref{fig:CIVnonBALQvBALQ} shows the \CIV\ EW versus the \CIV \ blueshift (hereafter the \CIV\ emission space) for the samples of non-BAL and BAL quasars. Let us first consider the non-BAL quasar population: consistent with \citet{Richards2011UnificationEmission}, weak lines (low EW) are often highly blueshifted, providing strong evidence of outflows, while strong lines are close to symmetric (zero blueshifts). As \citet{Richards2011UnificationEmission} point out, however, the relationship is not one-to-one as there are quasars with low blueshifts and low EWs. Note also that non-BAL refers here to quasars with both AI$=$0 and BI$=$0, i.e., no absorption present. If, however, the AI$>$0 quasars (still BI$=$0) are included, the non-BAL \CIV\ distribution does not in fact change significantly. Composite non-BAL quasars constructed from spectra in three regions of \CIV\ emission space are presented in the top panels of \autoref{fig:bal_nbal_compos}. \changemarkerii{In detail, a high-EW composite is generated from spectra with \CIV\ EWs and blueshifts of 103-116\,\AA\ and 262--787\kms, respectively; a low-EW, low-blueshift composite from 23--29\,\AA\ and -262--262\kms; and a low-EW, high-blueshift composite from 23--29\,\AA\ and 2360--2890\kms.} The spectra are smoothed using inverse-variance-weighting over a seven pixel window. The close to constant pixel-by-pixel median absolute deviation (MAD), shown as the shaded regions, redward of \CIV\ is almost entirely due to the finite S/N of the spectra\footnote{Spectrum S/N$\simeq$6, increased by $\sqrt{7}$ from the smoothing. Then, with MAD=$\sigma$/1.5, the amplitude of MAD/reconstruction is $\simeq\pm$0.04 in \autoref{fig:bal_nbal_compos}}. Such composite spectra, covering the full occupied region of the \CIV\ emission space, are available to view in an interactive plot online\footnote{\label{foot:url}\url{https://people.ast.cam.ac.uk/~alr53/BALs_nonBALs/CIV_BAL_nonBAL.html}}. 

The full spectrum reconstructions for the BAL-quasar population allow a direct comparison of the non-BAL- and BAL-quasar populations within the \CIV\ emission space (\autoref{fig:CIVnonBALQvBALQ}) 
and the bottom panels of \autoref{fig:bal_nbal_compos} contain composite BAL-quasars taken from the same areas in the \CIV\ emission space as the non-BAL quasar composites above. The same \CIV\ emission line shapes are apparent as for the non-BAL quasars. The BAL-quasar composites show systematically different BAL-trough properties moving around the \CIV\ emission space (see Section~\ref{sec:BALt}). Again, an interactive plot is available for the BAL quasars$^{\ref{foot:url}}$. \changemarkerii{The increase in the pixel-by-pixel MAD (shaded regions) within the troughs highlights the diversity in the absorber properties at a given velocity. The diversity of troughs is particularly evident in spectra with weak and blueshifted \CIV\ emission (lower right region of \autoref{fig:CIVnonBALQvBALQ}), an observation made in an important early paper discussing BAL-quasar properties by \citet{Turnshek1988BALSystems}.} We find that averaging over spectra possessing relatively-deep individual troughs at different velocities results in the composites showing weak but broad BAL troughs. The kinematics and properties of BAL troughs are explored more fully in Section~\ref{sec:BALt}.

\begin{figure}
    \centering
    \includegraphics[width=\linewidth]{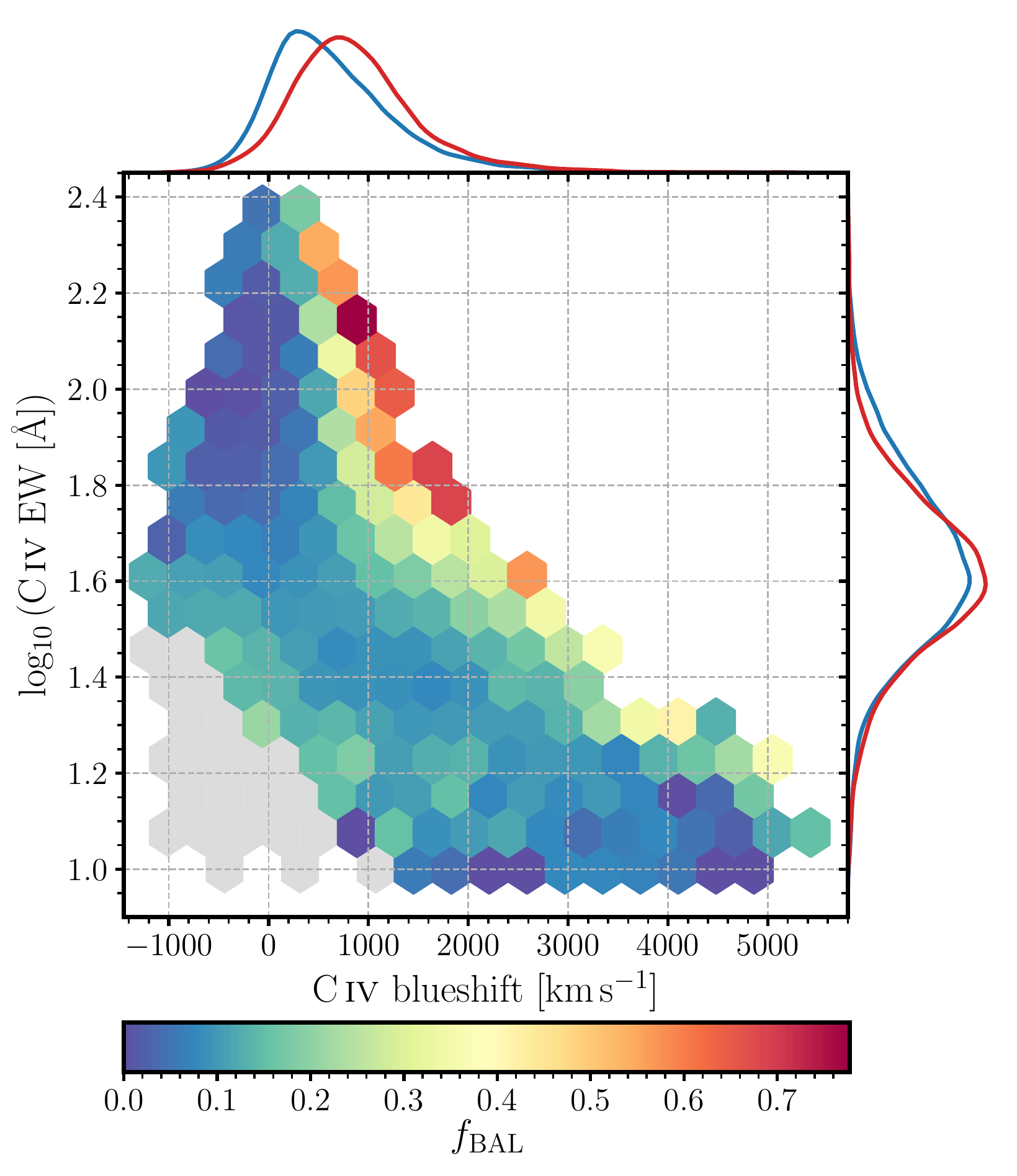}
    \caption{The observed fraction of quasars that are classically-defined BAL quasars as a function of the \CIV\ emission properties. Only bins with ten or more quasars are plotted. The grey hexagons indicate the sparsely populated region of \CIV\ emission space excluded from analysis. As discussed in the text, the intrinsic fraction of BALs is not known, however, at fixed EW, the probability a quasar possesses BAL troughs increases with increasing blueshift suggesting that the fraction of quasars with signatures of outflows in absorption increases with increasing evidence of outflows in emission. \changemarkerii{It is true that the fraction of quasars that are BALs, at a particular location in \CIV\ emission space, is highest at high-EW and positive-blueshift; however, these quasars have low BIs and the relative number of BAL quasars here is low compared to at higher \CIV\ blueshifts and lower EWs, as highlighted by the marginal distributions (same as for \autoref{fig:CIVnonBALQvBALQ}).}
    }
    \label{fig:bal_frac}
\end{figure}

As discussed in Section~\ref{sec:sample} the probability that a BAL-quasar is included in the DR14 catalogue as a function of SED, magnitude and redshift is not available. The level of uncertainty in the spectrophotometry of the DR14 quasar spectra also makes consideration of the amount of reddening experienced by the BAL-quasars far from straightforward. As a consequence, the quantitative assessment of the intrinsic frequency of BAL-quasars as a function of their BAL-trough properties and reddening is not possible \citep[cf.][particularly figure 22]{Allen2011AFraction}. Nonetheless, it is instructive to review whether the observed fraction of BAL-quasars in the sample changes across the \CIV\ emission parameter space.
\noindent
 $f_\text{BAL}$ is defined as 
\begin{equation}
    f_\text{BAL} = \frac{N_\text{BAL}}{N_\text{BAL}+N_\text{non-BAL}}.
\end{equation}
$N_\text{BAL}$ and $N_\text{non-BAL}$ depend on the measure used to select BAL quasars. Here we choose to employ the classic BI$>$0 definition to give the BAL-quasar sample, while the non-BALs are selected to possess both BI=0 and AI=0. \autoref{fig:bal_frac} shows $f_\text{BAL}$ over the \CIV\ emission space that includes both BAL and non-BAL quasars\footnote{\changemarker{The colour coding is essential to assimilate the information in \autoref{fig:bal_frac} and many subsequent figures. Readers are recommended to access the online journal if they do not have access to hardcopy version in colour.}}. For an extended range in \CIV\ emission EW, at constant EW, $f_\text{BAL}$ increases as the \CIV\ emission blueshift increases. If instead we consider the fraction of quasars with \textit{any} absorption greater than 450\kms\ wide and below \changemarkerii{0.95 times the continuum limit}, the previous statement still holds true but now there are quasars with absorption populating the high-EW, zero-blueshift region of \CIV\ emission space. For the reasons discussed above, caution should be taken when interpreting the quantitative behaviour of $f_\text{BAL}$ but the behaviour is consistent with an increased probability a quasar with a given \CIV\ EW is seen as a BAL as the \CIV\ emission outflow signature increases. \changemarkerii{This empirical observation may be further consistent with the theoretical wind model of \citet{Giustini2019AContext}, where the probability of a quasar hosting a radiation line driven wind increases with increasing accretion rate and BH mass.
}

\subsection
[C IV and and other UV emission lines]
{C\,{\sevensize IV} and other UV emission lines}
\label{sec:UVlines}

\begin{figure*}
    \includegraphics[width=\linewidth]{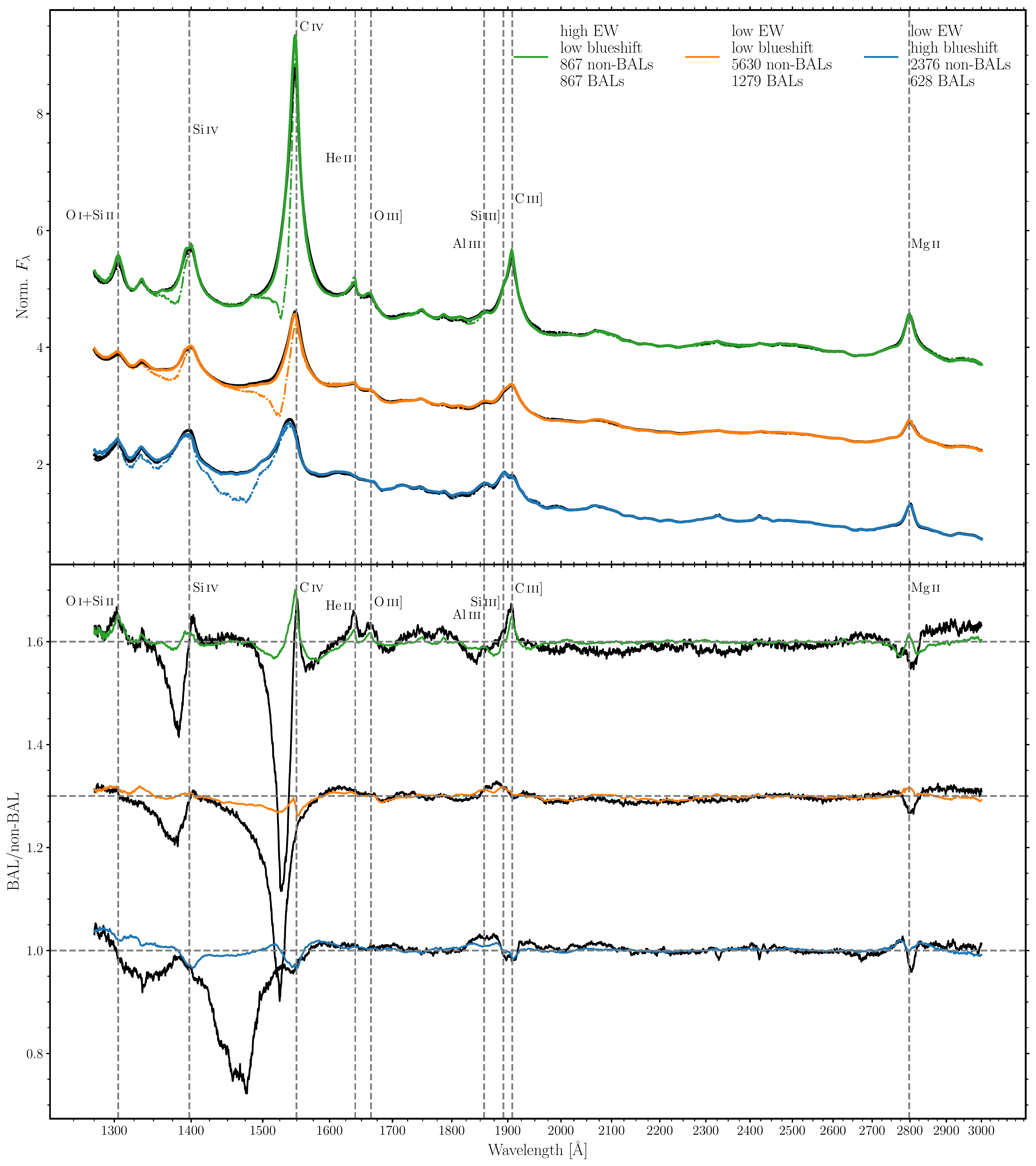}
    \caption{Composite spectra and reconstructions of the quasars in and surrounding the three EW and blueshift regions used to produce \autoref{fig:bal_nbal_compos} (top panel). The size of the areas in EW and blueshift have been extended to increase the number of quasars contributing to each composite. The BAL quasars follow the colours in the legend whilst the non-BAL quasars are plotted in black. The composite BAL (dot-dashed, coloured lines) and non-BAL (dot-dashed, black lines) quasar spectra are also plotted. Almost identical composites and reconstructions, such that the dot-dashed spectra are hardly visible underneath the reconstructions, mean the reconstructions are capturing the pertinent features in the spectra. The high-EW, low-blueshift composites have been shifted up by 1.5 and the low-EW, low-blueshift composite shifted up by 3 for presentation purposes. The bottom panel contains the BAL quasar composite spectra (reconstructions) divided by the non-BAL quasar composite spectra (reconstructions) in black (legend colours). Note that the y-scale has been chosen to emphasise even small differences between spectra (reconstructions). The high-EW composite and low-EW, low-blueshift composite have been shifted up by 0.6 and 0.3, respectively. Identical BAL and non-BAL quasar composite spectra (reconstructions) would produce divided spectra (reconstructions) of unity.}
    \label{fig:recon_div}
\end{figure*}

The composite spectra allow us to compare the BAL quasars to the non-BAL quasars in different regions of \CIV\ emission space. \autoref{fig:recon_div} shows the composites in \autoref{fig:bal_nbal_compos} over the wavelength range 1260--3000\,\AA. The BAL and non-BAL composite reconstructions in the top panel of \autoref{fig:recon_div} are extremely similar across the whole wavelength range. The composite spectra themselves are also very similar, apart from the presence of troughs in the BAL spectra, in agreement with \citet{Weymann1991ComparisonsObjects} and \citet{Matthews2017QuasarUnification}. There are, however, subtle differences in some of the emission lines at certain areas in \CIV\ emission space which can only be seen when dividing the BAL spectra (reconstructions) by the non-BAL spectra (reconstructions) as shown in the bottom panel. Note that the y-scale in the bottom panel has been chosen to illustrate low-amplitude spectral differences. 

Consider first the direct empirical differences evident in the BAL-quasar/non-BAL spectrum ratios. The BAL-absorption blueward of the \CIV \ and \SiIV \ emission is most evident as expected. All three composites also show a systematic difference in \MgII$\lambda$2800 emission with the BAL possessing weaker emission, although the reduction is small. The difference has been seen in other investigations \citep[see figure 8 in][for example]{Baskin2015OnQuasars} but is particularly evident here. Otherwise, the two composites drawn from the lower \CIV(EW) regions show no spectral differences greater than $\simeq$2 per cent shortward of the \CIV \ emission line.

\changemarkerii{The composite from the high \CIV(EW) and low \CIV \ blueshift region does, however, possess features coincident with wavelengths of prominent emission lines: peaks are present in the BAL-quasar/non-BAL spectrum ratio at }
\ion{O}{I}+\ion{Si}{II}, \SiIV, \CIV, \HeII, \OIII\ and \CIII. The location of quasars within the \CIV\ emission space means that the \CIV(EW)s are the same and care has been taken to ensure that both the distributions of \CIV(EW) and blueshift contributing to the non-BAL and BAL composites are carefully matched. The peak differences in emission-line strength are not large, a maximum of 10 per cent, but are highly significant. Inspection of the ratio spectrum reveals systematic reductions in the BAL-spectrum to the blue and red of several of the emission-line core excesses. The BAL quasar emission lines are \changemarkerii{narrower, in FWHM,} than those of the non-BAL quasars by $\simeq$14 per cent. The origin of the emission-line differences are therefore due to the BAL spectra possessing slightly narrower emission-line profiles than the non-BAL spectra. As far as we are aware the observation represents a new result. \changemarkerii{In the low-EW, high-blueshift composites, the converse is true but with less significance: the BAL quasar \CIV\ emission lines are wider in FWHM by $\simeq$7 per cent.}

Moving to the BAL/non-BAL reconstruction ratio spectra, the effectiveness of the reconstruction scheme can be assessed via comparison with the features evident in the BAL/non-BAL spectrum ratios\changemarkerii{, in particular the differences in emission lines between the two populations}. If the reconstructions were perfect the ratio spectra would essentially mimic the features seen in the spectrum ratios, other than at wavelengths affected by BAL troughs. Such is very definitely not the case across the \MgII$\lambda$2800 line. The reason, however, is due to the Baldwin effect, i.e., the anti-correlation between emission line EW and luminosity \citep{Baldwin1977LuminosityObjects}, which we have chosen not to remove. In more detail, the MFICA components were generated using quasars with redshifts $z=2.0-2.3$ (Section~\ref{sec:fitting}), thereby defining the relative emission-line strengths for the ultraviolet and \MgII \ lines. The quasars with the wavelength coverage of \MgII \ emission that contribute to the composites in \autoref{fig:recon_div}, however, have a significantly lower mean redshift, hence lower luminosity and thus higher emission-line EWs. The reconstruction emission-line strength for each quasar is dominated by the \CIV \ emission and other lines in the far ultraviolet. The MFICA-components possess the flexibility to reproduce the extended range of emission-line EW at any redshift but, statistically, the higher emission-line dominated MFICA-component contributions to the lower-redshift objects result in an over-prediction of the \MgII \ emission. For the two low-EW composites in \autoref{fig:recon_div}, the BAL and non-BAL ultraviolet emission has the same strength and the reconstructions across the \MgII \ line are identical. For the high-EW composite, the BAL ultraviolet emission is stronger by 10 per cent and the BAL quasar \MgII \ emission is thus also stronger, producing the visible `peak' at the location of the core of the \MgII \ line in the reconstruction ratio.

The Baldwin-effect induced mismatch to the \MgII \ emission aside the reconstruction-ratio spectra closely follow the spectrum ratios closely. Away from the \SiIV \ and \CIV \ BAL-troughs the reconstructions are within 2 per cent of the actual composite spectra. The reproduction of the individual emission-excesses for the BAL quasars in the top composite is striking. The similarities between the composite ratio spectra and the composite ratio reconstructions for all three regions in the \CIV\ emission space highlight the quality of our reconstruction scheme.

\begin{figure*}
    \centering
    \includegraphics[width=\linewidth]{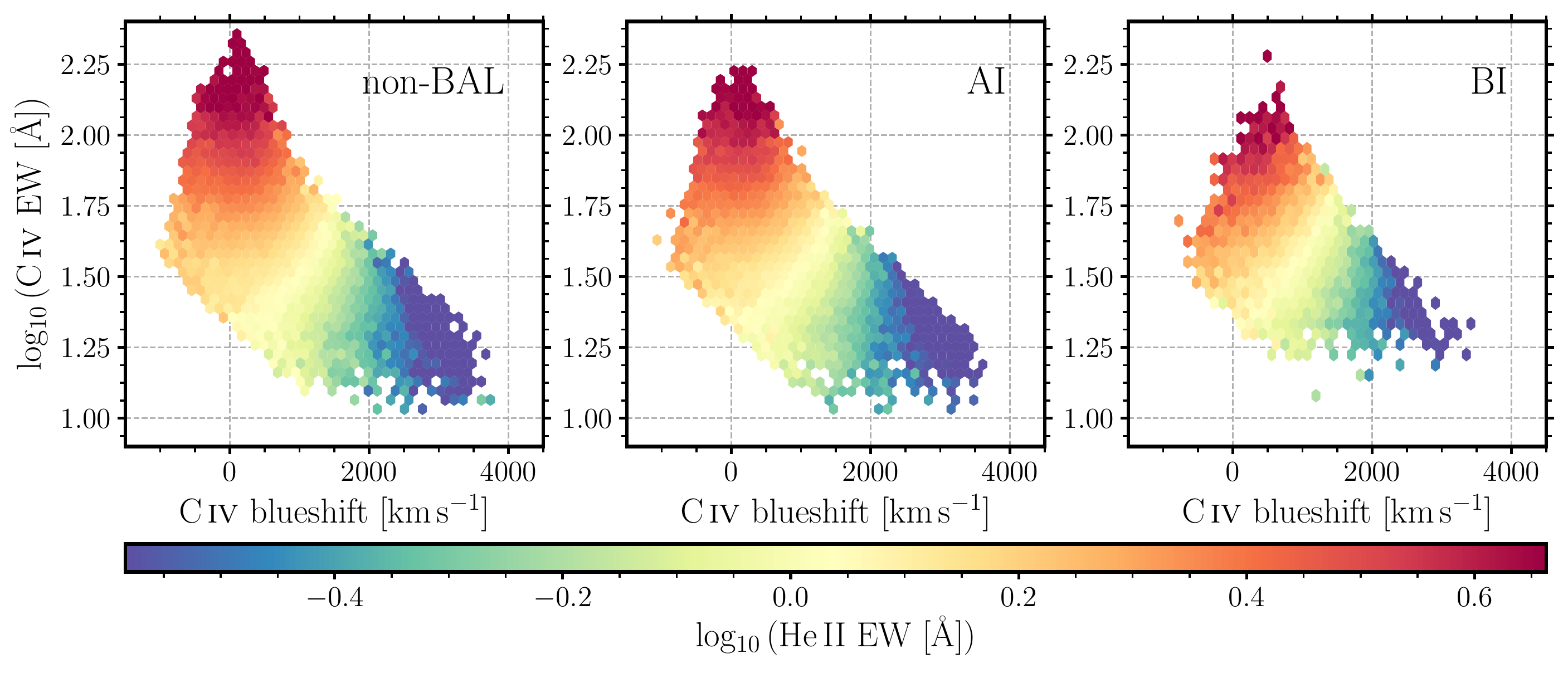}
    \caption{\CIV\ emission space with hexagons coloured by median \HeII\ EW for the non-BAL quasars (left), AI-defined quasars (middle; black histogram in \autoref{fig:AIpops}) and BI-defined BAL quasars (right; red, hatched histogram in \autoref{fig:AIpops}). Only hexagons with $\geq5$ quasars are plotted. The spectra in the lower left of \CIV\ emission space that were excluded and previously plotted as grey hexagons (\autoref{fig:bal_frac}) are not plotted here. The lower \CIV\ coverage in the BI-sample is a result of fewer quasars in this population. The \HeII-EW distributions are extremely similar in all three populations.}
    \label{fig:HeII}
\end{figure*}

High \HeII\ EW is indicative of a stronger soft X-ray spectrum \citep{Leighly2004HubbleInterpretation}. \citet{Baskin2013TheParameters, Baskin2015OnQuasars} noted the decrease in \HeII$\lambda$1640 strength with increasing \CIV\ blueshift by means of BAL and non-BAL quasar composites. With our reconstructions, we have been able to consider the relationship between the \CIV\ and \HeII\ on an object-by-object basis.
We calculate the \HeII$\lambda$1640 EW between 1620 and 1650\,\AA\ \citep[identical to][]{Baskin2013TheParameters} following the same procedure as for the \CIV\ doublet but using the windows 1610--1620\,\AA\ and 1700-1705\,\AA\ for the power-law estimation. Our observations are in agreement with \citet{Baskin2013TheParameters, Baskin2015OnQuasars} and the hypothesis that disc winds form only when the ionizing quasar SED is soft enough that electrons remain bound to nuclei and radiation line-driving contributes to the acceleration of material. The dependence of \HeII\ EW as a function of location in \CIV\ emission space confirms the anti-correlation between the hardness of the ionizing SED and the strength of outflows seen in both absorption and emission. A new result is the demonstration that the systematic \HeII \ EW trends for the non-BAL, AI- and BI-defined BAL quasar populations (\autoref{fig:HeII}) are close to identical.

\subsection{Black hole mass and luminosities}

\begin{figure*}
    \centering
    \includegraphics[width=\linewidth]{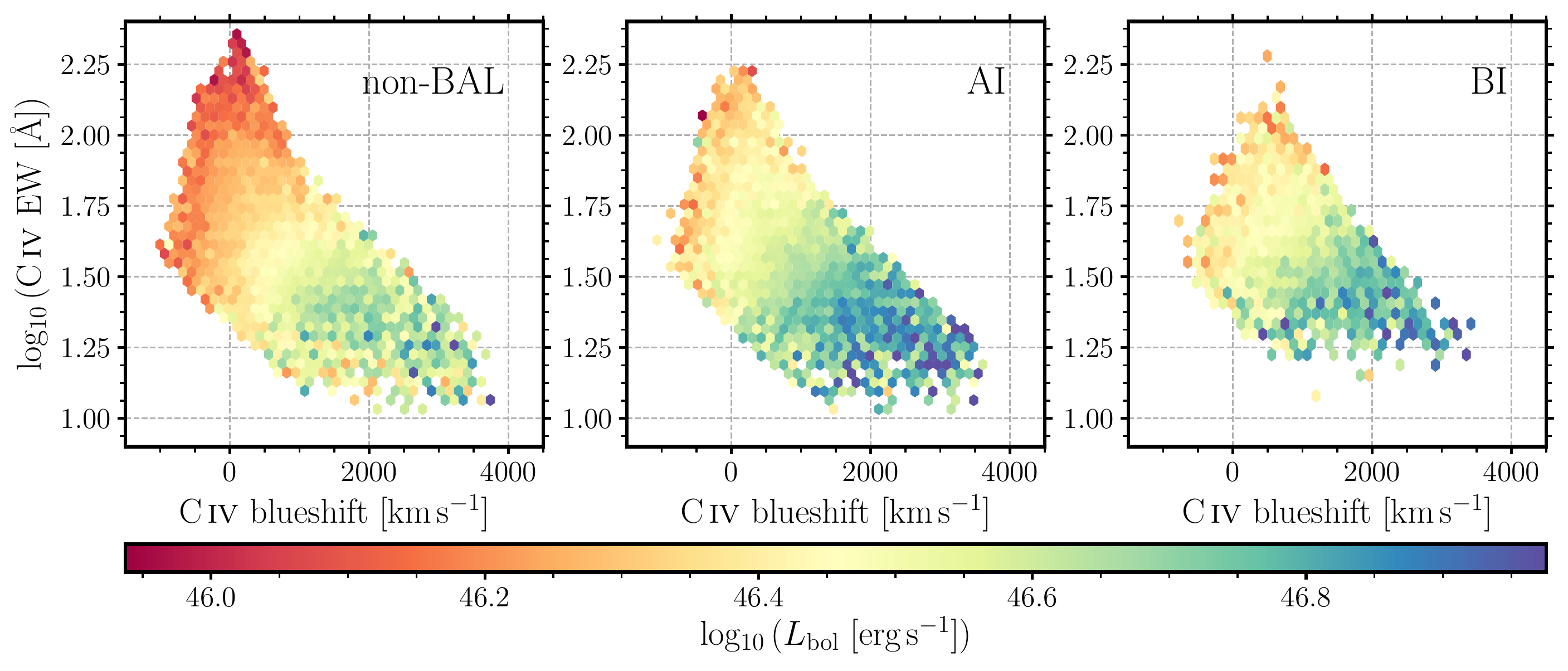}
    \caption{As for \autoref{fig:HeII} with hexagons occupied by five or more quasars coloured by median $L_{\text{bol}}$. Similar systemic trends within the \CIV\ emission space are present in all three populations.}
    \label{fig:Lbol}
\end{figure*}

The bolometric luminosity, $L_\text{bol}$, has been calculated from the monochromatic luminosity at 3000\,\AA\ or at 1350\,\AA\footnote{Spectrophotometric variations in the DR14 spectra contribute a $\pm$10 per cent uncertainty to the luminosity measurements but the factor is small relative to the dynamic range present in the quasar sample.} for spectra where 3000\,\AA\ is not present, using the bolometric corrections BC$_{3000}=5.15$ and BC$_{1350}=3.81$ from \citet{Shen2011A7}. $L_\text{bol}$ is displayed in \autoref{fig:Lbol} as a function of location in \CIV\ emission space for the non-BAL, AI- and BI-defined quasar populations. In agreement with the Baldwin Effect \citep{Baldwin1977LuminosityObjects}, albeit over a smaller dynamic range of luminosity, there is an anti-correlation between $L_\text{bol}$ and \CIV\ EW at fixed blueshift. $L_\text{bol}$ also increases towards the high-blueshift end of the \CIV\ emission space in all three of the populations. The three populations have very similar luminosities (median $\log_{10}(L_\text{bol})$=46.39, 46.57 and 46.54 for the non-BAL, AI and BI quasars respectively). Consideration of the intrinsic luminosities of the populations would require accurate extinction estimates for the quasars which are not available. It is, however, likely that the quasars with absorbers possess modest $E(B-V)$ values \citep[see][section 8.5]{Allen2011AFraction}. \changemarkerii{Adopting an average $E(B-V)=0.05$ for BAL-quasars \citep[from][]{Allen2011AFraction}, BAL-quasars are predicted to appear 25 per cent fainter at 3000\,\AA\ and 46 per cent fainter at 1350\,\AA. The reference-wavelength transition occurring at $z\simeq2.3$ when 3000\,\AA\ rest-frame moves beyond the red limit of the SDSS spectra. The intrinsic luminosities of the BAL-quasar samples should thus be larger by $\simeq$0.12 ($z\leq2.3$) and 0.27 ($z\geq2.3$) in $\log_{10}(L_\text{bol})$. The factors are modest relative to the dynamic range in luminosity present in the quasar samples and should not 
affect the trends within the \CIV\ emission space. Again, the systematic trends within the \CIV\ emission space for all three populations closely mimic each other.}

Only \CIV$\lambda$1550-emission is present in all the quasar spectra, thus we derive back hole mass estimates from the \CIV\ line and employ the correction of \citet{Coatman2017CorrectingMasses} to account for the excess, non-virial, blue emission for quasars with \CIV\ blueshift $>500$\kms. The \citet{Coatman2017CorrectingMasses} correction is not well-defined for modest positive blueshifts or negative blueshifts of any size. As a consequence, the mass-correction has not been applied where the \CIV\ blueshift $<500$\kms.

\begin{figure*}
    \centering
    \includegraphics[width=\linewidth]{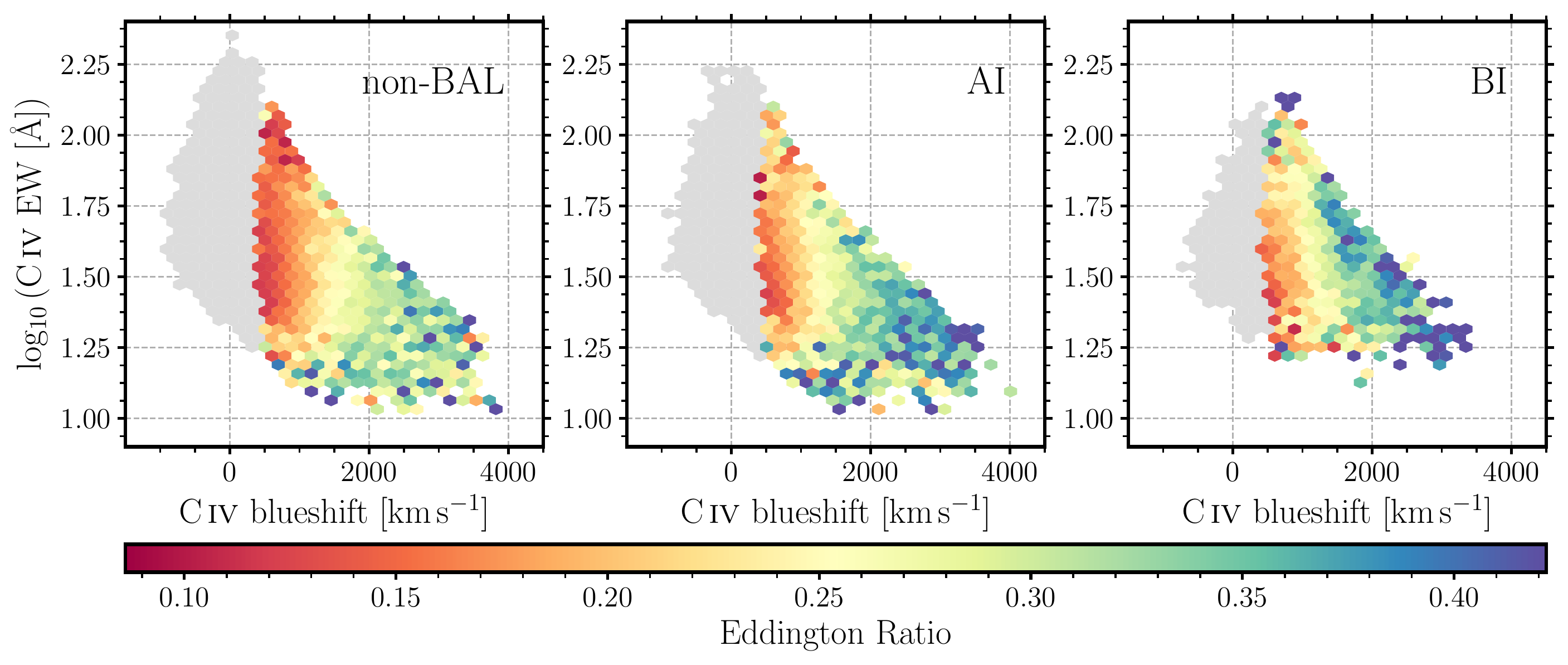}
    \caption{As for \autoref{fig:HeII} with hexagons occupied by five or more quasars coloured by median Eddington ratio. We only plot the Eddington ratio for spectra with \CIV\ blueshift $>500$\kms\ for which we correct the black hole masses using \citet{Coatman2017CorrectingMasses}. The Eddington ratio increases with \CIV\ blueshift for all three populations but is slightly lower at fixed \CIV\ parameters for the non-BAL population compared to the AI- and BI-defined samples (see also \autoref{fig:Lbol}).}
    \label{fig:Eddf}
\end{figure*}

From the black hole mass estimates we compute the Eddington luminosity, $L_\text{Edd}$, and thus the Eddington ratio using $L_\text{bol}$ from above, which we present as a function of \CIV\ emission space in \autoref{fig:Eddf}. Since the Eddington ratio depends on the Eddington luminosity which, in turn, depends on the black hole mass, we have only plotted the Eddington ratio for spectra with \CIV\ blueshift $>500$\kms. The Eddington ratios of the non-BAL, AI- and BI-defined quasars are similarly distributed in \CIV\ emission space, indicating that quasars with evidence for the strongest outflowing winds based on their emission-line properties have the highest $L/L_{\rm Edd}$.

\subsection{BAL trough parameters}
\label{sec:BALt}
\begin{figure}
    \centering
    \includegraphics[width=\linewidth]{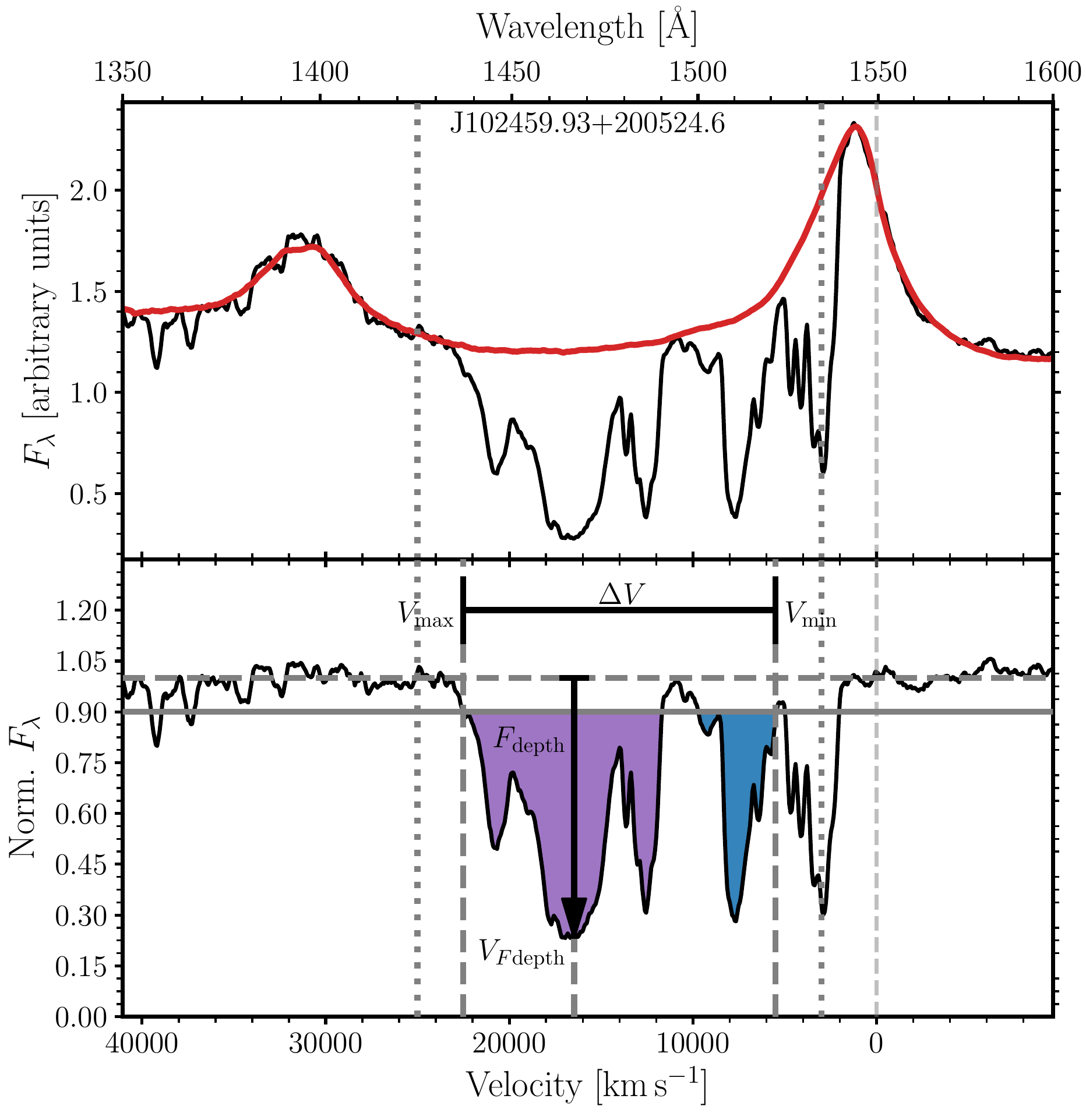}
    \caption{Example BAL quasar spectrum and reconstruction (top panel) alongside the normalised spectrum (bottom panel) in velocity space, inverse-variance weighted with a window of five pixels. The grey, dotted vertical lines mark the minimum and maximum velocities considered in the calculation of BI and thus the trough parameters. The dashed, grey line at zero velocity marks the laboratory wavelength of \CIV. The measured trough parameters for this spectrum are marked in the lower panel (see \autoref{tab:troughs} for descriptions of each parameter). There are two troughs in this BAL quasar spectrum (shaded regions) with the lower velocity, blue trough up against the minimum allowed velocity, although significant absorption at velocities below 3000\kms exists.}
    \label{fig:baldef}
\end{figure}

In \autoref{fig:bal_nbal_compos} and the interactive plot$^{\ref{foot:url}}$, we see the systematic trend in BAL trough properties with location in \CIV\ emission space. Strong and symmetric \CIV\ is often accompanied by relatively narrow and deep troughs while spectra with weak- and blueshifted-emission have broader and more diverse troughs. To investigate any systematic relationships between absorber-trough and emission-line properties, various trough parameters are measured. Each trough is defined as a region where the BI (or AI) conditions are met. \autoref{tab:troughs} lists the parameters and they are illustrated using an example spectrum in \autoref{fig:baldef}. Where spectra have multiple troughs, the parameters pertaining to individual trough measurements are reported as follows: we report the minimum $V_{\text{min}}$ (i.e., the lowest velocity for which there is broad absorption); the maximum $V_{\text{max}}$ (i.e., the highest velocity for which there is broad absorption); the deepest trough $F_{\text{depth}}$ measurement and the accompanying $V_{F\text{depth}}$, \changemarkerii{and the widest trough $V_{\text{width}}$.}

\begin{table*}
\caption{BAL trough parameters. All parameters with the subscript 450 pertain to troughs wider than 450\kms\ (e.g., $N_{450}$). Matching parameters also exist, with the subscript 2000, for troughs wider than 2000\kms, e.g., $N_{2000}$.}
\label{tab:troughs}
\begin{threeparttable}
\begin{tabular}{lcc}
\hline
Name & Unit & Description\\
\hline
BI & \kms\ & Balnicity index (\autoref{eqn:BI})\\
AI & \kms\ & Absorption index (\autoref{eqn:AI})\\
$N_{450}$ & & Number of troughs wider than 450\kms\\
$V_{\text{min},450}$ & \kms & Minimum velocity of each trough wider than 450\kms\\
$V_{\text{max},450}$ & \kms & Maximum velocity of each trough wider than 450\kms\\
$\Delta V_{450}$ & \kms & Range of velocities with absorption \\ 
$V_{\text{width},450}$ & \kms & Width of each trough \\
$F_{\text{depth},450}$ & Normalised flux units & Trough depth of each trough wider than 450\kms\\
$V_{F\text{depth},450}$ & \kms & Velocity of deepest part of each trough wider than 450\kms\\
\hline
\end{tabular}
\end{threeparttable}
\end{table*}

In \autoref{fig:CIVtrough1} we present the AI and BI trough parameters as a function of location in \CIV\ emission space. Considering the two populations of AI$>$0 quasars (\autoref{fig:AIpops}), we split the AI trough parameters into the two populations: AI troughs in spectra with BI$=$0 (\AIonly, left panels; blue histogram of \autoref{fig:AIpops}), and AI parameters for spectra with BI$>$0 (\AIBI, middle panels; orange histogram of \autoref{fig:AIpops}). The right-hand panels contain the BI trough parameters. By definition, almost all of the spectra with BI$>$0 have AI$>$0 and so appear in the middle and right panels\footnote{There is, however, a small fraction of the BI troughs ($\simeq$4 per cent) that have AI$=$0 due to $\chi^2_{\text{trough}}<10$ while there is no $\chi^2_{\text{trough}}$ requirement for the BI troughs.}. The extent in velocity of the absorption troughs therefore increases systematically moving from left to right in \autoref{fig:CIVtrough1}.

From first glance at \autoref{fig:CIVtrough1}, the  \AIonly\ and \AIBI\ trough systematics differ, while the latter are almost identical to the BI trough systematics, reinforcing the evidence for the existence of two AI populations described by \citet{Knigge2008TheQuasars} and presented in \autoref{fig:AIpops}. For spectra in the middle panels with only one AI trough, the AI trough parameters are identical to the BI trough parameters in the right panels except where $V_{\text{min},450}<3000$\kms. It is also possible for there to be more than one AI trough per spectrum in the middle panels; however, only one of these AI troughs has to satisfy the BI conditions for the object to be included in the right panels, thus the AI and BI trough parameters will differ. Approximately 54 per cent of the spectra in the middle panel have more than one trough.

In the top row of panels we see the dependence of absorption-outflow strength within the \CIV\ emission space, via the AI and BI measures. Clear differences exist between the \AIonly\ and the BI troughs. While the BI troughs become stronger with increasing \CIV\ emission blueshift, the \AIonly\ troughs show this same trend at fixed high-EW but at low-EW the AI measures decreases for large \CIV\ blueshifts. 

The left panel of the second row, showing the dependence of trough depth explains the \AIonly\ behaviour in the top row -- i.e., at \changemarkerii{$\log_{10}$\CIV(EW) $\lesssim1.6$} the troughs become shallower as \CIV\ blueshift increases thus AI also decreases (\autoref{eqn:AI}). The opposite is true at high \CIV\ EW. If there is any such systematic present for the BI-defined troughs it is far less significant, in contrast to what is seen in the composite troughs in \autoref{fig:bal_nbal_compos}. The weakening of the troughs in the composite spectra is a result of averaging out the individual troughs which are more diverse in velocity structure at higher \CIV\ blueshift and lower EW. The comparable depths of the troughs is clear in \autoref{fig:comp_trough} which shows absorber-restframe composites of the BI-defined BAL spectra in different regions of \CIV\ emission space. \citet{Baskin2015OnQuasars} presented similar absorber-frame composites and found trough depth not to be correlated with \HeII-EW (see their fig. 7), in agreement with what we present here in \CIV\ emission space.

The trends evident in the third, fourth and fifth rows, all showing measures of the absorber-trough velocities are however far more similar. For both AI-defined and BI-defined troughs the minimum and maximum velocities where absorption is present and the velocity of maximum absorption-depth all show strong systematics as a function of \CIV\ blueshift. As the strength of the emission-outflow signature increases, the velocities of the absorbers increase. Very few objects have $V_{F\text{depth}}$ at very large velocities (even at high \CIV\ blueshift) -- i.e., the deepest part of each trough often occurs in the lower-velocity half of the trough -- and this can also be seen in \autoref{fig:comp_trough}. \citet[see their fig. 4]{Hamann2019OnOutflows} observed these skewed absorption profiles in composites binned by BI. For both AI- and BI-defined samples, the highest-velocity troughs are in quasars with the highest Eddington ratio as observed in \autoref{fig:Eddf}.

The sixth and seventh row, examining the behaviour of the velocity-extent ($\Delta V$) and -width ($V_\text{width}$) of all and individual absorbers, respectively, within the spectra, provide insight into the systematic differences between the \AIonly\ and BI-defined absorbers. For the BI-defined absorbers as the \CIV\ emission blueshift increases, the trough velocities increase and the trough velocity-extent increases, as would be expected for an accelerating wind. The BAL troughs present in spectra in the lower right of \CIV\ emission space are much broader than those in spectra at the upper left. Whereas, for the \AIonly\ absorbers, the velocities increase as the \CIV\ blueshift increases but the velocity-extent of the absorbers is close to constant. 16\,390 AI-defined quasars have multiple troughs and in many they are separated extensively in velocity, up to 24\,950\kms. \changemarkerii{In the \AIonly, \AIBI\ and BI quasar populations, 67, 46 and 83 per cent of the quasars, respectively, have only one trough. Spectra with more than one trough typically have large trough velocities to allow for a second trough at lower velocities and thus have large \CIV\ blueshifts. Many of the \AIBI-quasars have a BI-defined trough (therefore also AI-defined) and a second AI-defined trough, accounting for the high fraction of spectra with $N_{450}>1$.}

\begin{figure*}
\includegraphics[width=\linewidth]{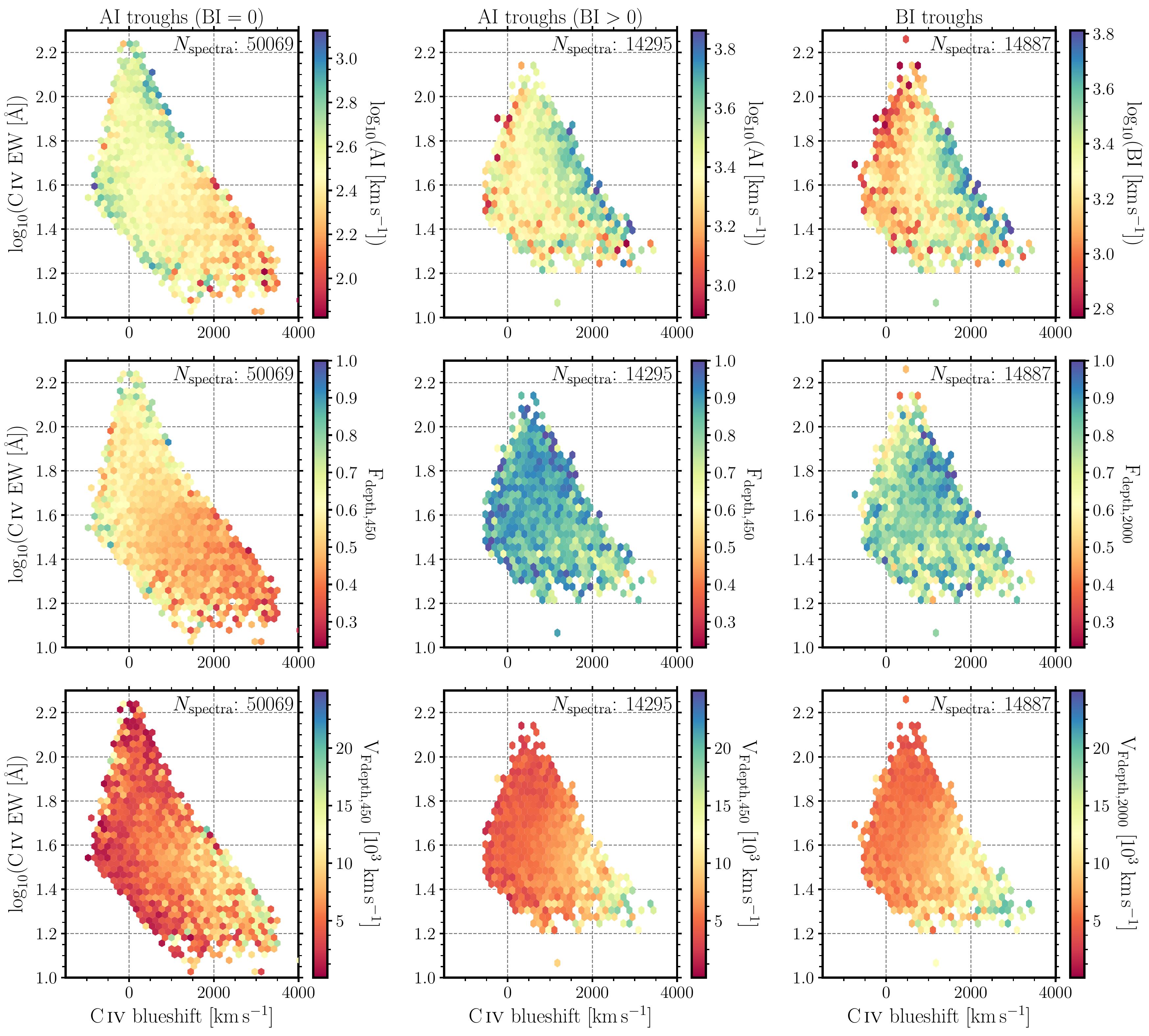}
\caption{The AI- and/or BI-defined quasar populations binned in \CIV\ emission profile space where each hexagonal bin is coloured by trough parameters outlined in \autoref{tab:troughs}. Left panels: AI trough parameters for spectra with BI$=$0; middle: AI trough parameters for spectra with BI$>$0; right: BI trough parameters for classically-defined BAL quasars. The quasars have been binned in \CIV\ emission space and the median trough parameter in each bin plotted on the colour axis, for bins with at least five quasars. The number of spectra contributing to each panel is noted in the top right of every panel; each row has the same number of spectra. From top to bottom, AI or BI, $F_{\text{depth}}$ and $V_{F\text{depth}}$. The figure continues overleaf with $V_{\text{min}}$, $V_{\text{max}}$, $\Delta V$ and $V_{\text{width}}$. The dot-dashed lines in the lower right panel locate the boundaries in \CIV\ emission space used to produce the absorber-restframe composites in \autoref{fig:comp_trough}. The structure of both the AI- and BI-defined troughs vary systematically, but differently, in \CIV\ emission space.}
\label{fig:CIVtrough1}
\end{figure*}

The \AIonly\ spectra are dominated by single, relatively narrow, troughs such that BI$=$0. These narrow troughs appear in spectra with the full range in \CIV\ blueshift as evidenced by the swathe of constant and low AI, $\Delta V_{450}$ and $V_{\text{width},450}$ in the \AIonly\ plots. We can test the hypothesis that the \AIonly\ absorbers are different from the \AIBI\ absorbers rather than just much narrower by investigating the structure of the zero-BI troughs. \changemarkerii{We select 18\,740 \AIonly\ quasars} with only one AI-trough and generate a mean composite of the spectra shifted to the rest-frame of the maximum absorption-depth. A well-defined \CIV\ doublet is observed (\autoref{fig:CIVdoublet}). In many cases, the 450\kms\ minimum velocity required by the AI measurement is sufficiently narrow to include narrow absorption lines (NALs) with very different internal kinematic properties compared to BI-defined absorbers. The AI-defined absorption is participating in the outflows but we conclude that \AIonly\ absorbers are dominated by material in structures with velocity spreads of $\simeq$few-hundred\kms, very different from much of the material giving rise to the classical BI-defined troughs.

We have yet to consider the high-EW, low-blueshift ($\lesssim$1000\kms) region of \CIV\ emission space. Inspection of the $\Delta V$ distribution for the \AIonly\ troughs shows increased $\Delta V$ at high \CIV\ EW. Some of these troughs have $V_{\text{width},450}>2000$\kms\ but have BI$=$0 as most of the absorption is below 3000\kms. In contrast with the NALs discussed above, as the \CIV\ blueshift increases, over an admittedly small range of velocities present at high EW, AI and $V_{\text{width},450}$ increase whilst the troughs deepen ($F_{\text{depth},450}$ increases). \changemarkerii{In other words, while the minimum velocity remains constant, the maximum velocity increases. The original BI calculation was designed with the 3000\kms\ starting velocity to exclude strong associated absorbers which are believed to result from a different phenomenon compared to the BAL troughs \citep{Weymann1991ComparisonsObjects}. In the circumstances, the use of a BI$_0$ definition for absorbers (where the BI metric is extended down to zero velocity) will include a larger fraction of absorbers resulting from outflows but at the expense of including true `associated absorbers'.}

\begin{figure*}
\includegraphics[width=\linewidth]{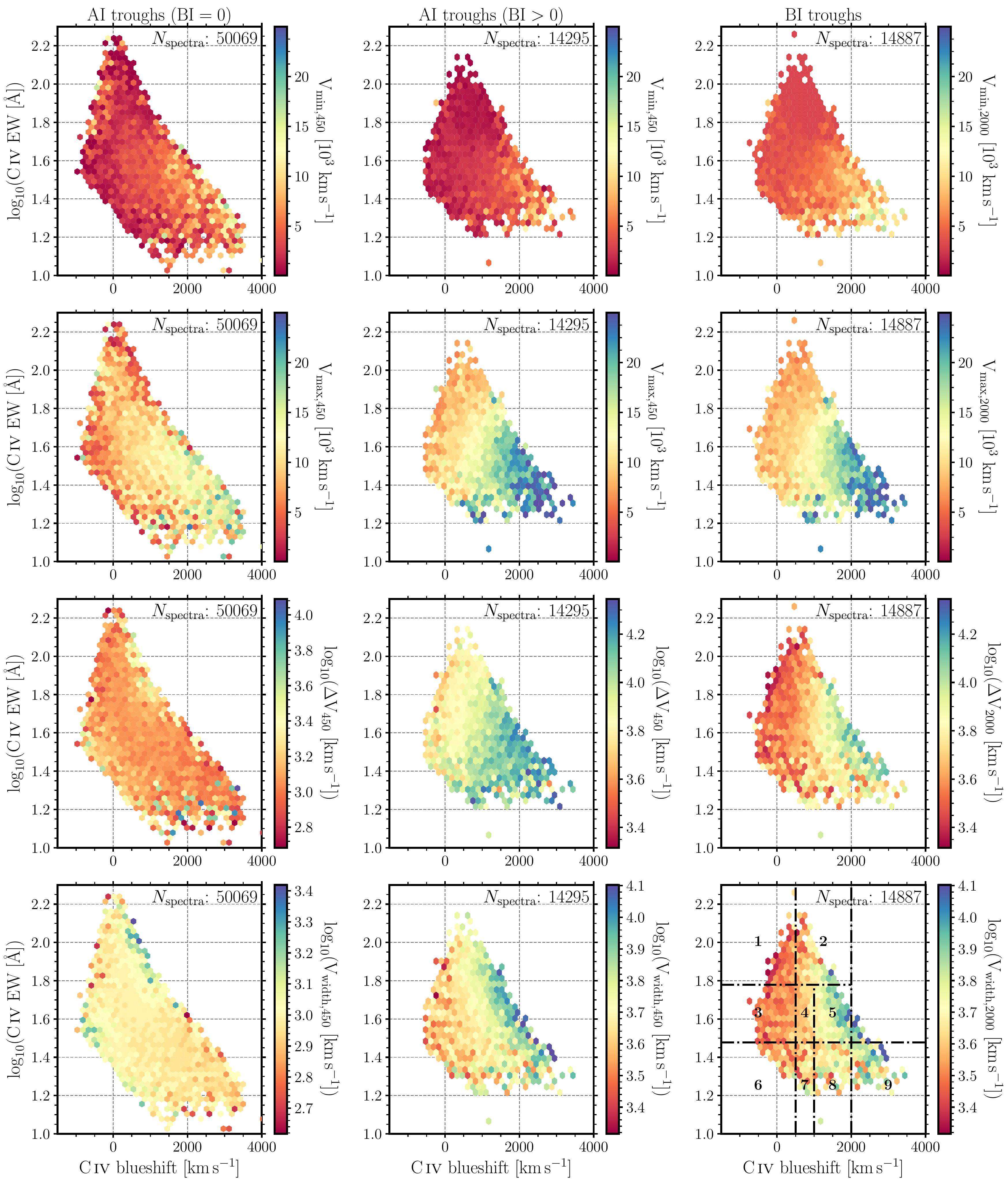}
\contcaption{}
\label{fig:CIVtrough2}
\end{figure*}

\begin{figure}
    \centering
    \includegraphics[width=\linewidth]{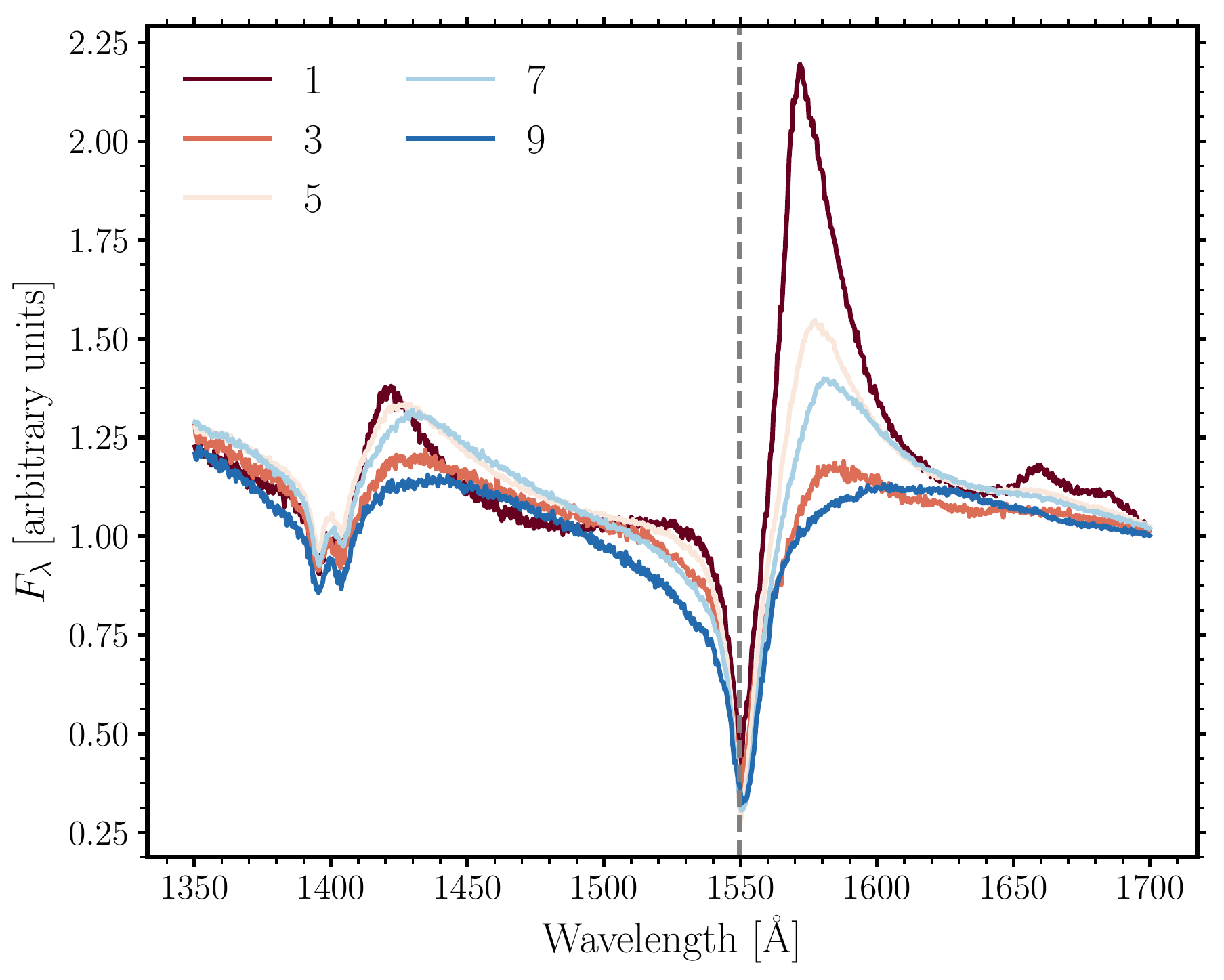}
    \caption{Median absorber-restframe composites of the 14\,887 BI-defined quasar spectra. Where spectra have more than one trough, the deepest trough is selected. \changemarkerii{The numbering of the composites matches that of the \CIV\ regions in the lower right panel of \autoref{fig:CIVtrough1}; however only plotting the odd-numbered composites for clarity.} The BI-defined troughs extend to larger velocities at higher \CIV\ blueshifts with the maximum velocity increasing more rapidly than the minimum velocity or the velocity of the deepest part of the trough. The depth of the troughs does not vary systematically in \CIV\ emission space.}
    \label{fig:comp_trough}
\end{figure}

\begin{figure}
    \centering
    \includegraphics[width=\linewidth]{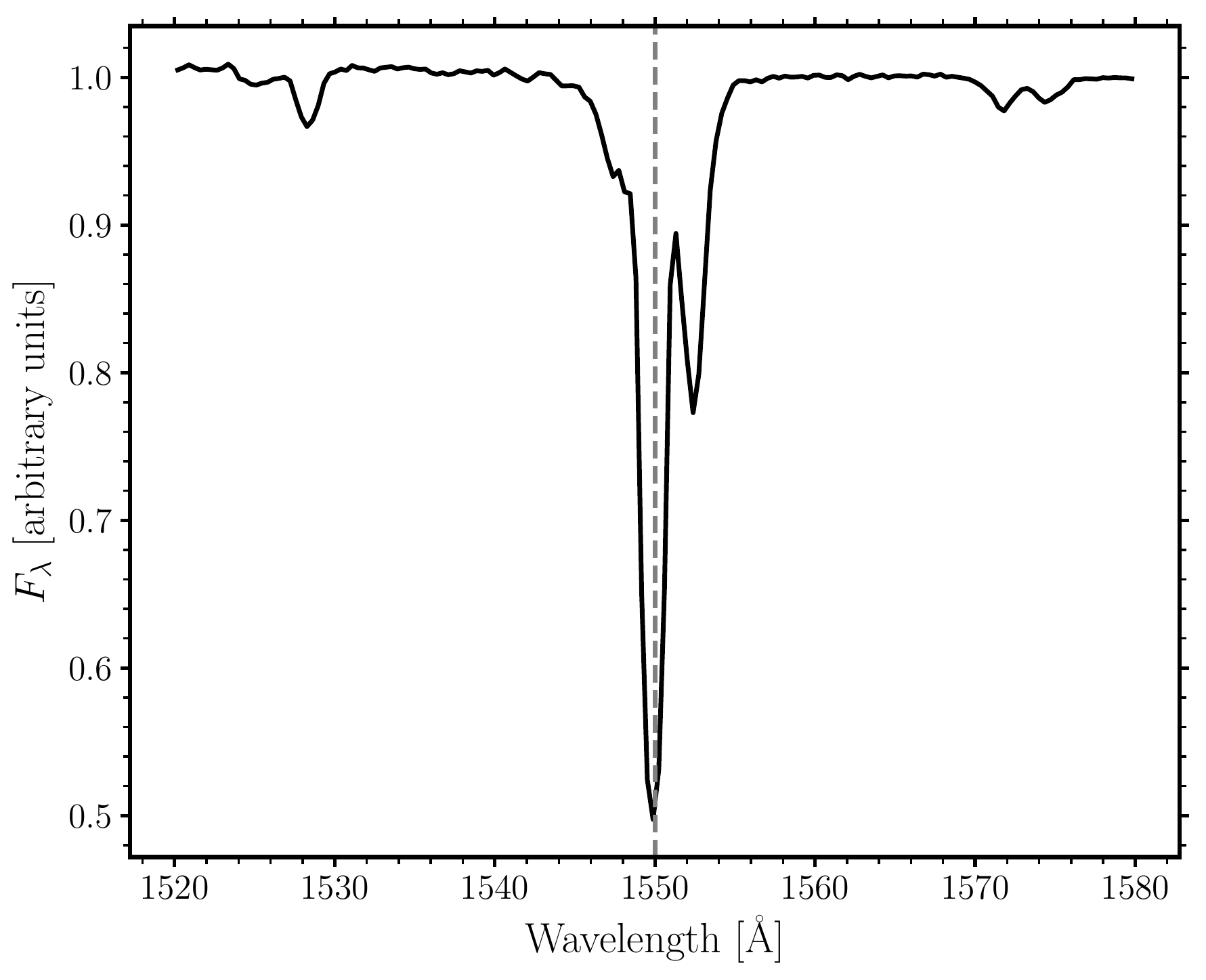}
    \caption{\changemarkerii{Mean normalised composite of 18\,740 single, \AIonly\ troughs shifted to the rest-frame of the deepest part of the trough. The composite is dominated by the well-defined deep, narrow absorption in which both features in the \CIV$\lambda\lambda$1548,1551 doublet are seen. Additional weaker absorption is present in some troughs, explaining the broader depression, particularly to shorter wavelengths.}}
    \label{fig:CIVdoublet}
\end{figure}

\section{Discussion}
\label{sec:disc}
The main aim of this paper is to present the observational differences and similarities between the ultraviolet spectra of non-BAL and BAL quasars, including the outflow signatures evident in the \CIV$\lambda$1550 emission. Our approach has been designed to allow a direct investigation of the systematics observed in the absorption- and emission-outflow properties for the non-BAL and high-ionisation BAL populations as a whole. In short, we can definitively locate BAL quasars in \CIV\ emission-line parameter space, even in the face of significantly absorbed \CIV\ profiles. The results should provide constraints on models, particularly in the case of disc winds, but we do not attempt to propose a quantitative model, deferring such considerations to a future paper.

\subsection{Quasar physical properties}
\label{sec:physdisc}

The observational manifestation of outflows observed in absorption and emission are quite different, particularly in terms of the solid angle over which material contributes to the spectrum. BAL- and narrow-absorption occurs due to gas present along a particular line of sight. For absorbing structures of limited physical extent, a small change in viewing angle may result in very different observed absorber properties. Similar changes may result temporally due to rapid motion of absorbing material across the line of sight or changes in the ionization parameter. Emission-line properties instead derive from gas covering a much larger solid angle from the perspective of the observer. Photons produced by emission, in gas participating in an outflow, over a significant angle contribute to the observed spectrum and small changes in viewing angle are not, in general, expected to result in significant changes to the emission-line spectrum. Observed emission-line changes due to larger angular variations in viewing angle are of course potentially powerful probes of non-spherical geometries, including disc-wind models.

A main result is the similarity in the observed ultraviolet-emission and physical properties between the BAL and non-BAL quasar populations. For every BAL quasar with a certain black hole mass, bolometric luminosity, Eddington ratio, \CIV\ emission profile and \HeII$\lambda$1640 EW, there exists a non-BAL quasar with essentially identical properties, consistent with the two objects possessing the same orientation and/or geometry. \changemarkerii{\citet{Matthews2017QuasarUnification} suggest that BAL quasars can be observed at similar inclinations to non-BAL quasars.} The converse is not always true; while for the majority of the \CIV\ emission space one can find a BAL-quasar for each non-BAL quasar of specified properties, there is a lack of BAL-quasars possessing large \CIV\ emission EW, no significant \CIV\ emission blueshifts and hard SEDs.

\citet{Richards2011UnificationEmission} used the observed relation between the morphology of the \CIII$\lambda$1908-\SiIII$\lambda$1892-\AlIII$\lambda$1857 complex and the blueshift of the \CIV\ emission\footnote{A similar approach has been adopted here (see Appendix~\ref{app:priors}) to determine the priors adopted for the reconstruction of the \CIV\ emission profiles.} to deduce the locations of BAL-quasars within the \CIV\ emission space. The conclusions of \citet{Richards2011UnificationEmission} are consistent with the results presented here but now far more information is available about the location of BAL-quasars in the \CIV\ emission space as a function of the BAL-trough properties (see \autoref{fig:CIVtrough1} in particular). 

The distribution of quasars in \CIV\ emission space (\autoref{fig:CIVnonBALQvBALQ}) can be explained by a disc-wind model \citep[e.g.,][]{Richards2011UnificationEmission}; the Doppler blueshifted emission is thought to occur in an outflowing wind component, thus spectra with symmetric and high-EW \CIV\ show little to no evidence of a disc-wind and instead are dominated by disc emission. Quasars with extremely blueshifted emission are all weak-lined. Evidence from the strength of \CIV\ and \HeII-emission (see below) indicates that quasars at the top-left of the \CIV\ emission space have the hardest SEDs, with a systematic softening of the ionizing continuum moving towards the bottom right. 
In a scenario where line-driving is important for producing the wind, only when the number of high-energy ionizing-photons is small enough that electrons remain bound to nuclei can the wind be accelerated to high velocities. As a consequence, quasars with strong \CIV\ emission \textit{and} high blueshift are not found.

As a recombination line, \HeII$\lambda$1640 can be used as an indicator of the number of ionizing-photons with energies above 54\,eV, i.e., as an SED-hardness diagnostic. In the line-driven wind scenario, only the spectra with the lowest \HeII\ EW, thus fewest high-energy ionizing-photons, are able to produce the strongest disc-winds. \citet{Baskin2013TheParameters, Baskin2015OnQuasars} have quantified the strength of the dependence of BAL-trough properties as a function of the \HeII-emission EW (broader and faster-moving troughs are observed when \HeII \ is weak). The systematic behaviour of \HeII-emission as a function of \CIV\ blueshift for non-BAL quasars has also been identified by \citet{Baskin2015OnQuasars}. Again, our results (\autoref{fig:HeII}) for the strength of \HeII-emission in the BAL and non-BAL populations across the \CIV\ emission space show striking systematic behaviour and demonstrate the very close similarity in properties for BAL and non-BAL populations. 

Simple comparisons of emission-line properties for the BAL and non-BAL populations as a whole will show differences as a result of the lack of BAL-quasars with hard SEDs and strong, symmetric \CIV$\lambda$1550 emission. The dearth of BAL quasars at the top left of the \CIV\ emission space (\autoref{fig:CIVnonBALQvBALQ}) has naturally led to the conclusion that BAL-quasars are more common among the quasar population with softer SEDs. \changemarkerii{Our analysis indicates the situation is somewhat more complex as BAL-quasars exist across the entire remainder of the \CIV\ emission space occupied by the non-BAL quasars. While SED-hardness is important, a stronger emission-outflow signature at fixed \CIV\ EW is also relevant for increasing the probability that a quasar is a BAL quasar 
(\autoref{fig:bal_frac})}. Overall, therefore, a key conclusion from the results presented in Section~\ref{sec:res}, is that, when an outflow is present, \changemarkerii{i.e., \CIV\ blueshift > 0,} the physical properties (including $L$, $L/L_{\rm Edd}$, SED-hardness) and ultraviolet emission-line properties of the non-BAL and BAL populations are essentially indistinguishable.

\subsection{Absorber properties}

\citet{Knigge2008TheQuasars} were the first to propose the existence of two-populations of absorbers among quasars with positive AI values. In our absorber classification a physically-significant difference is evident between the AI(BI=0))-defined absorbers and those defined by \AIBI. Row five of \autoref{fig:CIVtrough1} shows how, independent of the AI- or BI-value of troughs, the maximum absorber outflow velocity increases with increasing \CIV\ emission blueshift. Row six, however, reveals a clear difference in the velocity extent of troughs, with AI(BI=0)-defined absorbers possessing a constant velocity width ($\Delta V$), while the \AIBI\ absorbers show a strong systematic trend of increasing $\Delta V$ with increasing maximum outflow velocity. The latter behaviour is consistent with outflowing absorbing material in an accelerating wind. The composite absorption spectrum of the AI(BI=0)) absorbers (\autoref{fig:CIVdoublet}) by contrast shows both components of the \CIV$\lambda\lambda$1548,1551 doublet are visible. The typical velocity-spread within the absorber structures, independent of outflow velocity, must, therefore, be no more than $\simeq$250\kms.

It is in the high-EW region of \CIV\ emission space where quasars have the hardest ionizing-SEDs that we observe slightly narrower emission lines in the BAL-quasars than the non-BALs (\autoref{fig:recon_div}). If absorption is present in these spectra, it is typically deep and also narrower than the majority of the BI-defined troughs but is amongst the broader AI-defined troughs. The narrower emission lines and hard SEDs, combined with the differing trough systematics, could suggest a different orientation of the quasars or different outflow-driving mechanism \citep{Richards2012CIVWinds} from that of the outflows traced by the absorption seen in all other areas of \CIV\ emission space. Alternatively, certain sight-lines may probe different parts of the outflows, which may be dominated by different driving mechanisms. 

The investigation presented here has focussed on quasars with BAL-troughs while a study of the outflow properties of narrow \CIV\ absorption lines in non-BAL and BAL quasars was undertaken by \citet{Bowler2014Line-drivenQuasars}. Results included the detection of line-locked absorbers but, most relevant to findings presented here, also showed that the kinematics and population statistics of narrow absorbers in the non-BAL and BAL quasar sub-populations were similar. Combining the results with those presented here leads to a picture where, in respect of outflow signatures, the non-BAL and BAL quasar populations possess the same emission and NAL outflow properties strongly suggesting a direct link between apparent `non-BAL' and `BAL' quasars.

While the results in Section~\ref{sec:physdisc} support the idea that BAL quasars have the same parent population as non-BALs \citep{Richards2006AGNPerspective}, it is clear that the properties of the absorption troughs change significantly across the \CIV\ emission space, \changemarkerii{which is an observation that is expected from theory \citep[see][]{Giustini2019AContext}}. The observation may be important in the context of understanding the origin of `BAL quasars' as population statistics, such as the fraction of quasars classified as BALs, are potentially a strong function of the BAL definition. Put another way, BAL quasars can be found across nearly the full \CIV \ emission space but BAL-trough properties change in such a way that grouping all BALs together into the same subclass of objects may restrict the physical insight that can be gained into the origin and properties of outflows.

\section{Conclusions}
\label{sec:conclude}

Using mean-field independent component analysis, we have successfully reconstructed the ultraviolet spectra of both high-ionization BAL and non-BAL quasars from the SDSS DR14 quasar catalogue with redshifts $1.56\leq z\leq3.5$ and S/N $\geq5$; $\simeq$144\,000 quasar spectra in total (\autoref{fig:examp}). Using our reconstructions and employing improved redshifts (\autoref{fig:DR14_z}), we have redefined the BAL quasar population (\autoref{fig:DR14_BI}) and found evidence for two AI-defined populations, in agreement with \citet{Knigge2008TheQuasars}: BI$=$0 and BI$>$0 quasars  (\autoref{fig:AIpops}). 

By reconstructing the quasar spectra, we have recovered the intrinsic \CIV\ emission of BAL quasars even where said emission is heavily absorbed. This has allowed us to, for the first time, place the BAL quasars alongside the non-BAL quasars in \CIV\ emission profile space (\autoref{fig:CIVnonBALQvBALQ}). We find that for every BAL quasar with certain \CIV\ emission properties, there exists a non-BAL quasar with the same properties. \changemarkerii{The converse is not necessarily true; there are almost no BAL quasars at the highest EWs in \CIV\ emission space with zero blueshift, compared to the non-BAL quasars.}

In addition to \CIV\ emission, BAL and non-BAL quasars are extremely similar in respect of their \HeII$\lambda$1640 EW, bolometric luminosity, and Eddington ratio (Figs.~\ref{fig:HeII}--\ref{fig:Eddf}). The similarities in properties, some measured from the reconstructed emission spectra, suggest that the broad absorption observed in BAL quasars is the result of clumpy, outflowing gas along the line-of-sight \citep[broadly consistent with][]{Yong2018UsingStructure} and that all quasars -- barring perhaps those with the hardest ionizing-SEDs at the \changemarkerii{highest \CIV\ emission EWs and lowest blueshifts} -- have the potential to be seen as BAL quasars but with varying probability (\autoref{fig:bal_frac}). The trough parameters also vary systematically in \CIV\ emission space (\autoref{fig:CIVtrough1}), and we have gained insight into the trough-structure of the two AI-defined populations. 
At the highest \CIV\ EWs, the BAL quasars have slightly narrower emission lines compared to the non-BAL quasars (\autoref{fig:recon_div}) and it is also these BAL quasars that have quite different troughs from the rest of the BAL quasars -- deep, narrow and close to systemic velocity -- which perhaps could suggest a different driving mechanism of the outflows and/or a different quasar orientation.

Virtually every paper on BAL quasars starts by noting the two main hypotheses, which are that BALs are ubiquitous in the quasar population but have only a $\sim$20 per cent covering fraction or that they represent a distinct 20 per cent of the quasar population (with nearly 100 per cent covering fraction). With this paper we argue that the question is largely resolved (at least for high-ionization BALs): the similarity of BAL and non-BAL quasars demands that BALs are not a distinct class of quasars. Rather, BAL quasars are normal quasars observed along a particular line of sight or at a particular time. Conceivably, both may be true given that BAL troughs can be transient \citep[e.g.,][]{Capellupo2012VariabilityVariability, Rogerson2018EmergenceOutflows, Yi2019BroadQuasar}. 
Models invoking non-spherical geometries and particular viewing angles, e.g., BALs seen when the line-of-sight probes just above/below some form of obscuring torus, are not straightforward to reconcile with the strong systematic relationship between absorber and emission-line kinematics across the full range of \CIV\ emission properties.

\section*{Acknowledgements}
We thank Bob Carswell, Norm Murray and Vivienne Wild for helpful discussions. An anonymous referee provided a comprehensive review for which we are grateful. ALR acknowledges funding via the award of an STFC studentship. MB acknowledges funding from The Royal Society via a University Research Fellowship. PCH acknowledges funding from STFC via the Institute of Astronomy, Cambridge, Consolidated Grant. 

Funding for the Sloan Digital Sky Survey IV has been provided by the Alfred P. Sloan Foundation, the U.S. Department of Energy Office of Science, and the Participating Institutions. SDSS-IV acknowledges
support and resources from the Center for High-Performance Computing at
the University of Utah. The SDSS web site is www.sdss.org.

SDSS-IV is managed by the Astrophysical Research Consortium for the 
Participating Institutions of the SDSS Collaboration including the 
Brazilian Participation Group, the Carnegie Institution for Science, 
Carnegie Mellon University, the Chilean Participation Group, the French Participation Group, Harvard-Smithsonian Center for Astrophysics, 
Instituto de Astrof\'isica de Canarias, The Johns Hopkins University, Kavli Institute for the Physics and Mathematics of the Universe (IPMU) / 
University of Tokyo, the Korean Participation Group, Lawrence Berkeley National Laboratory, 
Leibniz Institut f\"ur Astrophysik Potsdam (AIP),  
Max-Planck-Institut f\"ur Astronomie (MPIA Heidelberg), 
Max-Planck-Institut f\"ur Astrophysik (MPA Garching), 
Max-Planck-Institut f\"ur Extraterrestrische Physik (MPE), 
National Astronomical Observatories of China, New Mexico State University, 
New York University, University of Notre Dame, 
Observat\'ario Nacional / MCTI, The Ohio State University, 
Pennsylvania State University, Shanghai Astronomical Observatory, 
United Kingdom Participation Group,
Universidad Nacional Aut\'onoma de M\'exico, University of Arizona, 
University of Colorado Boulder, University of Oxford, University of Portsmouth, 
University of Utah, University of Virginia, University of Washington, University of Wisconsin, 
Vanderbilt University, and Yale University.




\bibliographystyle{mnras}
\bibliography{main} 


\appendix
\section{Creation of priors}
\label{app:priors}
\begin{figure}
\includegraphics[width=\linewidth]{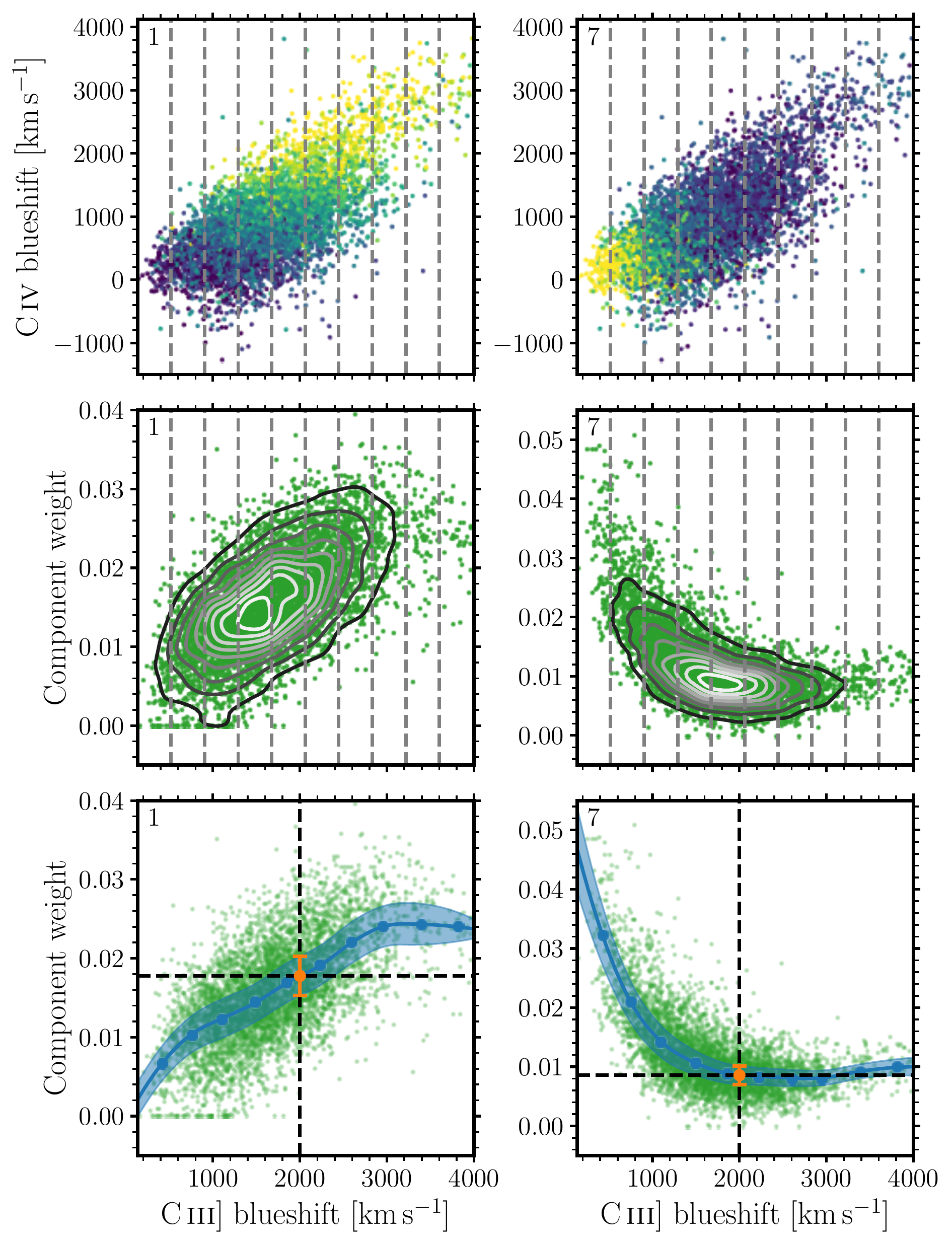}
\caption{Priors for first component (left panels) and 7th (right panels) as examples, created from $\sim$4000 non-BAL quasars. Top: \CIV\ blueshift against that of \CIII\ for the non-BAL quasars, with points coloured by component weight. Vertical grey lines mark the \CIII\ `blueshift' bin edges. Middle: component weight against \CIII\ `blueshift'. Grey lines are again the bin edges. Bottom: median component weight in each \CIII\ `blueshift' bin against \CIII\ `blueshift' (blue points). The blue line and shaded regions are created via a quadratic spline interpolation of the median and the median absolute deviation component weights. The green points are the same as those in the middle panels. \changemarkerii{The orange points mark the priors for an example spectrum with a \CIII\ blueshift of 2000\kms\ where the points and errorbars mark the centre and width of the normal priors for each component.}}
\label{fig:priors}
\end{figure}

The correlation between the morphology of the \CIII$\lambda$1908+\SiIII$\lambda$1892+\AlIII$\lambda$1857 emission complex and the extent of the blueshift of \CIV\ emission in luminous quasar spectra has been known for some time \citep[see fig. 16 of][and the top panels of our \autoref{fig:priors}]{Richards2011UnificationEmission}. While the \CIV\ blueshift is a result of shifting the centroid of the line, the observed \CIII\ `blueshift' is caused by an increase in the ratio of \SiIII\ to \CIII\ emission. \citet{Richards2011UnificationEmission} used the \CIII\ `blueshift' to parametrize the blueshift of the absorbed \CIV\ emission in a sample of BAL quasars. Here, we use the \CIII\ `blueshift' to place priors on the 10 component weights for the BAL quasars since, for the high-ionisation BAL-quasars under investigation, the 1600--2900\,\AA \ wavelength interval is free of high-ionisation transitions where broad absorption is observed.

\begin{figure}
\includegraphics[width=\linewidth]{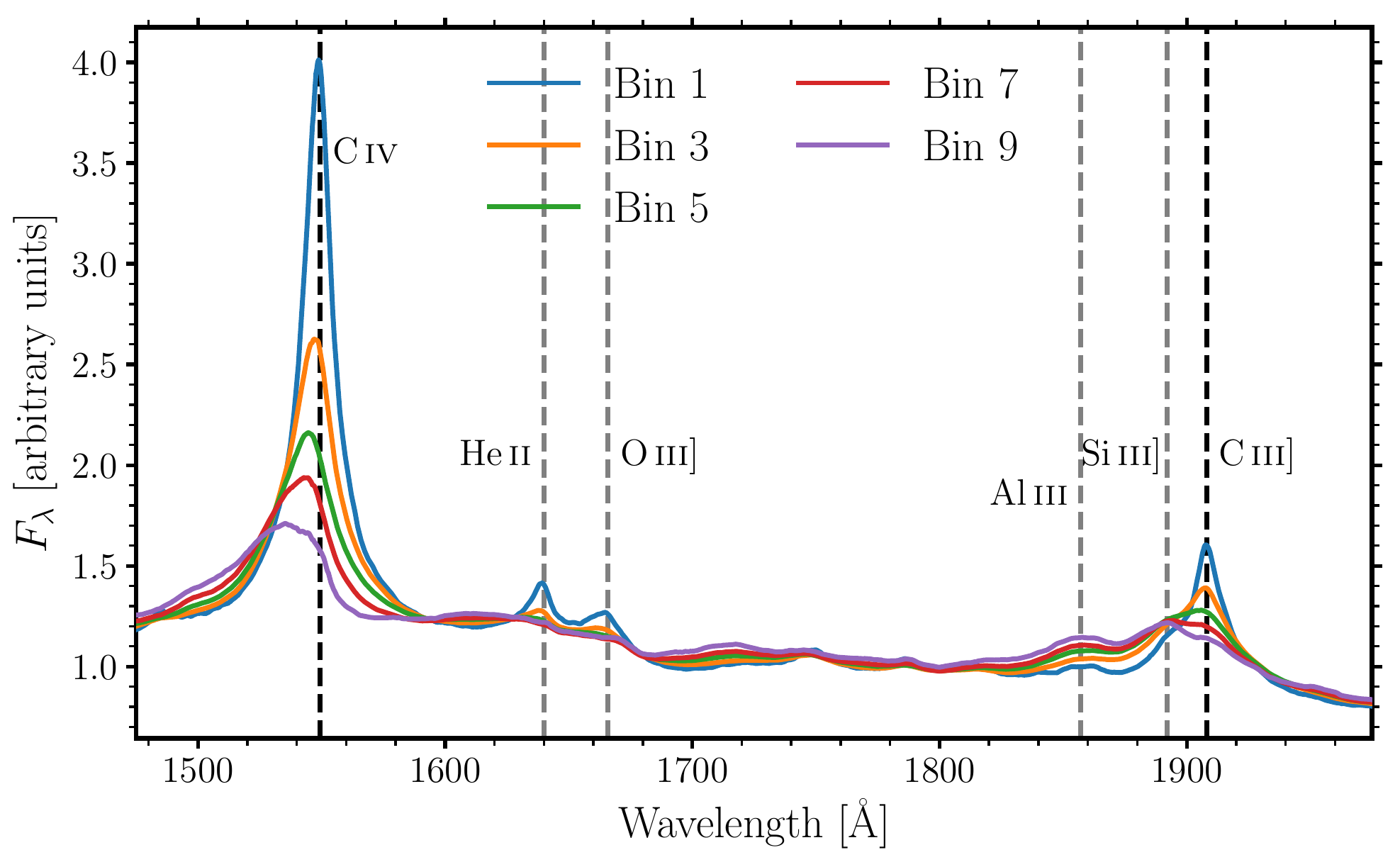}
\caption{Non-BAL quasar composite reconstructions for the odd-numbered bins in \CIII\ `blueshift'. From the first bin to the 9th, as the \CIII\ `blueshift' increases the \CIV\ blueshift also increases.}
\label{fig:pricomposrecon}
\end{figure}

The component-weight priors were constructed using a sample of $\sim$4000 non-BAL quasars with S/N $\geq9$ per pixel. Similarities between the BAL quasar and non-BAL quasar spectra in the wavelength range 1600--3000\,\AA\ allow the non-BAL quasars to be used to infer the BAL quasar component weights. The components covering 1260--3000\,\AA\ were fitted to the spectra and the \CIII\ `blueshift' measured between 1820 and 1970\,\AA\ using the recipe detailed in Section~\ref{sec:CIVem}. The continuum was fitted using the wavelength intervals 1790--1810\,\AA\ and 2015--2035\,\AA. The middle panels of \autoref{fig:priors} show two examples of the component weights against the \CIII\ `blueshift' for the non-BAL quasar sample. The spectra (and component weights) were binned according to the \CIII\ `blueshift'. The edges of the bins are drawn on the top four panels in \autoref{fig:priors}. The correlation between the \CIII\ and \CIV\ is evident in the composite reconstruction for each bin in \autoref{fig:pricomposrecon}: with increasing \CIII\ `blueshift' (increasing bin number), the \CIV\ blueshift also increases. Note also that the \CIV\ EW decreases and that \HeII$\lambda$1640 decreases in strength with increasing blueshift, the latter of which is in agreement with \citet{Richards2011UnificationEmission} and \citet{Baskin2013TheParameters, Baskin2015OnQuasars}.

The median and median absolute deviation (MAD) of each component weight in each bin was calculated (see examples in the bottom panels of \autoref{fig:priors}). From these measurements, for each \CIII\ `blueshift' measured in the whole quasar sample, normal priors can be created by implementing quadratic spline interpolation via the Python library SciPy (shaded regions in the same panels). As an example, a spectrum with a \CIII\ blueshift of 2000\kms\ will have a normal prior for component 1 (7) with centre 0.018 (0.009) and width 0.005 (0.003).

\section{The unabsorbed SEDs of BAL and non-BAL quasar populations}
\label{app:similar}
\begin{figure*}
\includegraphics[width=\linewidth]{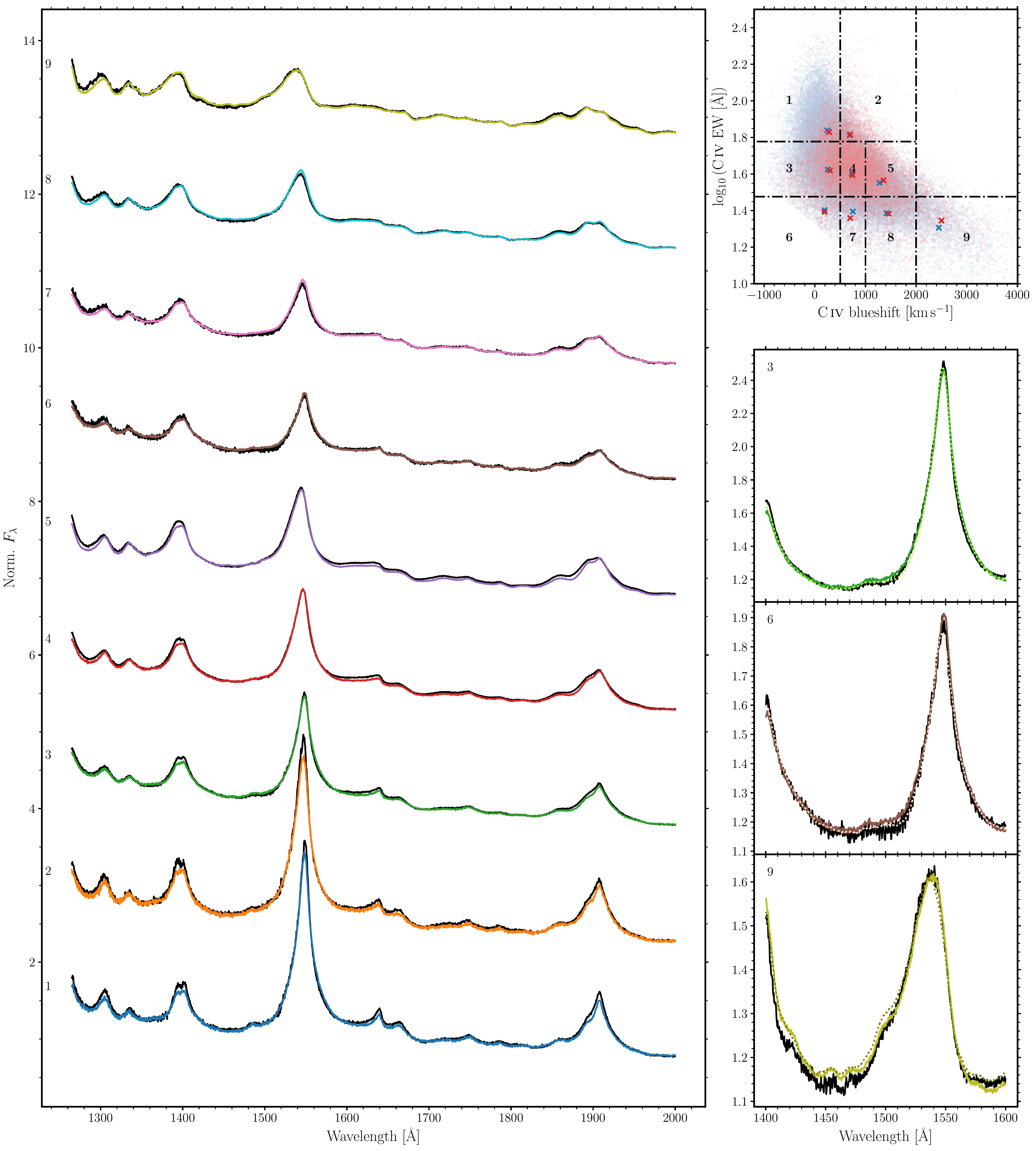}
\caption{Composite spectra from the numbered \CIV\ regions in the top right panel. BAL composite spectra are in black and non-BAL quasars in colours. Each composite is numbered according to its location in \CIV\ emission space. For three of the regions we plot the \CIV\ emission in the lower right panels. Composite BAL reconstructions are also plotted here (coloured dotted lines). The \CIV\ parameters measured from the composite spectra (crosses) \changemarkerii{and from individual quasars (points)} are plotted in the \CIV\ emission space with non-BALs in blue and BALs in red.}
\label{fig:CIVspectra}
\end{figure*}

All quasar spectra in our sample, whether BAL or non-BAL, are reconstructed using MFICA components generated from a sample of non-BAL quasars. Much of the BAL spectra blueward of 1600\,\AA\ is affected by absorption such that to reconstruct the BAL quasar spectra we rely on the assumption that the correlation between the \CIV\ emission and the spectra redward of 1600\,\AA\ of the non-BAL quasars can also be applied to the BAL quasar spectra. To test the validity of the assumption that, excluding absorption, the BAL and non-BAL spectra are similar, we have generated composite BAL and non-BAL spectra for different regions in \CIV\ emission space. The regions are those marked on the \CIV\ emission space in the top right panel of \autoref{fig:CIVspectra}. \changemarkerii{Pixels masked due to narrow or broad absorption} in individual spectra do not contribute to the composites. In regions 1--4, where the distributions of BAL and non-BAL quasars in the \CIV\ emission space differ significantly the objects contributing to each composite have been matched carefully. Specifically, non-BAL composites have been constructed using the same number of spectra contributing to the BAL composite, where each non-BAL spectrum has been selected to lie close to a BAL-spectrum in the \CIV\ emission space.

\autoref{fig:CIVspectra} presents the composite spectra of the BAL and non-BAL quasars, illustrating that for all regions in \CIV\ emission space the median non-BAL and BAL spectra are remarkably similar. Only by adopting a very contrived model could the results of the comparison (for the two populations) be explained in circumstances where the assumption was not also valid for individual BAL-quasars. There are only slight differences in a few regions, namely 3, 6 and 9, where the continuum blueward of \CIV\ is lower in the BAL composites than the non-BAL composites (1, 2 and 1 per cent difference on average for the three regions) on account of imperfect masking of the absorption troughs. The composite BAL \textit{reconstructions} (lower right panels of \autoref{fig:CIVspectra}) are almost identical to the non-BAL spectra. The improved match comes about because of the continuity over hundreds of \kms\ in the reconstruction process further reducing the effect of the modest depression in the BAL-spectra.

The \CIV\ emission properties computed from the BAL and non-BAL composite spectra and the BAL-reconstructions show only very small differences. Even using the composite spectra themselves, the differences in the \CIV\ EW and blueshift are small, typically only $\simeq$3\, per cent ifn EW and $\simeq$70\kms\ in blueshift. No blueshift difference exceeds 150\kms\ and the most extreme difference in EW occurs for the composites in region 9, where the 2\,\AA \ change corresponds to a 9\,per cent increase for the BAL composite. The measurements from the BAL and non-BAL reconstructions, used throughout the paper, are even more similar. 

The differences in the \CIV\ emission properties between the actual BAL and non-BAL composite spectra are negligible in the context of all measurements presented in the paper. The result provides clear evidence that the assumption regarding the similarity between the BAL and non-BAL SEDs is valid to a high degree of accuracy for the two populations. 



\bsp	
\label{lastpage}
\end{document}